\documentclass[preprint]{aastex}

\def\etal{{\it et~al.\,}}

\def\mj{$M_{\rm J}\,$}

\def\Dwa{$\,$\uppercase\expandafter{\romannumeral5}$\,$}

\def\sless{\lower2pt\hbox{$\buildrel {\scriptstyle <}
   \over {\scriptstyle\sim}$}}
\def\sgreat{\lower2pt\hbox{$\buildrel {\scriptstyle >}
   \over {\scriptstyle\sim}$}}
\def\aa{Astron. Astrophys.\ }
\def\apjl{Astrophys.~J.~Letters\ }
\slugcomment{Submitted to the Ap. J.}
\begin{document}
\title{Phase Functions and Light Curves of Wide Separation Extrasolar Giant Planets}
\author{David Sudarsky\altaffilmark{1}, Adam Burrows\altaffilmark{1},
Ivan Hubeny\altaffilmark{1}, \& Aigen Li\altaffilmark{1,2}}
\altaffiltext{1}{Department of Astronomy and Steward Observatory, 
                 The University of Arizona, Tucson, AZ \ 85721}
\altaffiltext{2}{Lunar and Planetary Laboratory, The University of Arizona,
                 Tucson, AZ \ 85721}
\begin{abstract}
We calculate self-consistent extrasolar giant planet (EGP) phase functions and
light curves for orbital distances ranging from 0.2 AU to 15 AU.  We explore the
dependence on wavelength, cloud condensation, and Keplerian orbital elements.
We find that the light curves of EGPs
depend strongly on wavelength, the presence
of clouds, and cloud particle sizes.  Furthermore, the optical and
infrared colors of most EGPs are phase-dependent, tending to be reddest at crescent phases
in $V-R$ and $R-I$.  Assuming circular orbits, we find that at optical wavelengths most EGPs are
3 to 4 times brighter near full phase than near greatest elongation for highly inclined
(i.e., close to edge-on) orbits.
Furthermore, we show that the planet/star flux ratios depend strongly on the Keplerian
elements of the orbit, particularly inclination and eccentricity.
Given a sufficiently eccentric orbit, an EGP's atmosphere may make periodic transitions from
cloudy to cloud-free, an effect that may be reflected in the shape and magnitude
of the planet's light curve.  Such elliptical orbits also introduce
an offset between the time of the planet's light curve maximum and the time of full
planetary phase, and for some sets of orbital parameters, this light curve maximum can be
a steeply increasing function of eccentricity.  We investigate the detectability of
EGPs by proposed space-based direct-imaging instruments.

\end{abstract}

\section{Introduction}
Over the past several years, a wide variety of nearby extrasolar giant planet
(EGP) systems has been detected indirectly by radial velocity techniques
(e.g., Mayor \& Queloz 1995; Marcy \& Butler 1996; Marcy \etal 1998, 1999, 2002;
Butler \etal 1997, 1999, 2002; Queloz \etal 2000).  Model atmospheres and spectra
for a fraction of these objects have been developed (Seager \& Sasselov 1998;
Barman \etal 2001; Sudarsky, Burrows, \& Hubeny 2003; Burrows, Sudarsky,
\& Hubeny 2004) to guide ground- and space-based
observations.  However, most studies to date have not addressed
the dependence of EGP spectra and colors on planetary phase, which, along
with ephemerides, may be crucial to the interpretation of directly imaged EGPs.

Some attention has been given to the self-consistent modeling
of planetary phase functions and light curves for the close-in
EGPs ($\lesssim$ 0.05 AU; Seager \etal 2000; Green \etal 2003), but
at wider separations, simple Lambert reflection models (isotropic surface
reflection; Sobolev 1975) or fits to reflection
data for the giant planets of our solar system have been utilized.  While
much of their recent paper focuses on ringed extrasolar planets,
Arnold \& Schneider (2004) study the phase dependence of a ringless planet
as a function of orbital inclination by assuming Lambert reflection.  The
spectral dependence of the light curves is not investigated, and a geometric
albedo is assumed, rather than computed.  Dyudina \etal (2004)
model the light curves of ringed and ringless
EGPs by using red-band {\it Pioneer} phase data for Jupiter
and Saturn.  A functional fit to these data allows them to construct planetary
phase functions, from new to full phase.  EGP light curves for various
distances and eccentricities are then derived, assuming these planetary
phase curves.  Due to the limitations of the {\it Pioneer} data, they could not
investigate the dependence on wavelength
or cloud condensates and particle sizes.  Furthermore, they could not explore
the dependence of the albedos and phase functions on orbital distance.

With numerous space-based instruments under development
that will be capable of detecting EGPs
in the optical or near infrared spectral regions
(e.g. {\it Eclipse}, Trauger \etal 2000, 2001;
{\it EPIC}, Clampin \etal 2002, Lyon \etal 2003;
{\it Terrestrial Planet Finder}, Beichman \etal 2002;
{\it Kepler}, Koch \etal 1998; {\it Corot}, Antonello \& Ruiz 2002; {\it MONS},
Christensen-Dalsgaard 2000; {\it MOST}, Matthews \etal 2001),
wavelength-dependent planetary phase models for a range of orbital
distances will be in demand.
In this work, we compute self-consistent, solar-metallicity atmosphere models of EGPs
and derive albedos, phase integrals, phase functions, and
light curves as a function of wavelength.  The effects of varying
the Keplerian elements, including semi-major axis, orbital inclination, eccentricity,
argument of periastron, and longitude of the ascending node are
investigated.  Because atmospheric compositions vary enormously
with orbital distance, a detailed and self-consistent approach is necessary.
Our present model set spans orbital radii from
0.2 AU to 15 AU, thereby encompassing a great variety of atmospheric
structures.

In \S\ref{sec_numerical}, we review the formalism associated with our treatment
of planetary phase functions and include a description of our numerical methods.
Section \ref{sec_solarphase} contains a sampling of Solar System phase functions,
while \S\ref{sec_EGPphase} details our model phase functions for EGPs.  Section \ref{sec_albedos}
contains geometric and spherical albedo spectra for a full range
of EGP orbital distances.  Wavelength-dependent EGP light curves for circular orbits 
are presented in \S\ref{sec_lightcurves}.  Section \ref{sec_colors} details the
dependence of EGP colors on planetary phase.  Section \ref{sec_partsizes}
investigates the effects of cloud particle size variation on EGP light curves.  
In \S\ref{sec_elliptical}, we explore the important effects of eccentricity
and viewing angle on the resulting light curves.  We conclude with a summary and discuss
in \S\ref{sec_summary} prospects for the detection of wide-separation EGPs and their light curves.

\section{Formalism and Numerical Techniques} 
\label{sec_numerical}
It is useful to review important definitions and quantities relevant to
the study of planetary phases and light curves.
Planetary phase is a function of the observer-planet-star orientation,
and the angle whose vertex lies at the planet is known as the {\it phase
angle} ($\alpha$).  The requisite formalism for the computation of
planetary brightness as a function of phase angle has been presented
by numerous authors.  Following Sobolev (1975), we relate the
planetary latitude ($\psi$) and longitude ($\xi$)
to the cosine of the angle of incident radiation ($\mu_0$) and the cosine
of the angle of emergent radiation ($\mu$) at each point on the planet's
surface: 
\begin{equation}
\mu_0 = \cos\psi\cos(\alpha-\xi)
\label{eq_latlong}
\end{equation}
and
\begin{equation}
\mu = \cos\psi\cos\xi,
\end{equation}
where latitude is measured from the orbital plane and longitude is measured
from the observer's line of sight.  The phase angle is then,
\begin{equation}
\alpha = \cos^{-1}\left(\mu\mu_0 - \left[{(1-\mu^2)(1-\mu_0^2)}\right]^{1/2}cos\phi\right),
\end{equation}
where $\phi$ is the azimuthal angle between the incident and
emergent radiation at a point on the planet's surface.  The emergent
intensity from a given planetary latitude and longitude
is given by
\begin{equation}
I(\mu,\mu_0,\phi) = \mu_0S\rho(\mu,\mu_0,\phi),
\end{equation}
where the incident flux on a small patch of the planet's surface
is $\pi\mu_0S$, and $\rho(\mu,\mu_0,\phi)$ is the reflection coefficient.
In order to compute the energy reflected off the entire planet, one must
integrate over the surface of the planet.  For a given planetary phase,
the energy per second per unit area per unit solid angle received by an observer is 
\begin{equation}
E(\alpha) = 2S{R_p^2\over{d^2}}\int_{\alpha-\pi/2}^{\pi/2}\cos(\alpha-\xi)
\cos(\xi)d\xi\int_0^{\pi/2}\rho(\mu,\mu_0,\phi)\cos^3\psi
d\psi,
\end{equation}
where $R_p$ is the planet's radius and $d$ is the distance to the
observer.  This quantity is related to the {\it geometric albedo} ($A_g$), the reflectivity
of an object at full phase ($\alpha=0$) relative to that of a perfect Lambert disk of
the same radius under the same incident flux, by
\begin{equation}
A_g = {{E_p(0)d^2}\over{\pi SR^2}}.
\end{equation}
A planet in orbit about its central star displays
a range of phases, and the planet/star flux ratio is given by
\begin{equation}
{F_p\over{F_*}} = A_g\left({R_p\over{a}}\right)^2\Phi(\alpha),
\end{equation}
where $\Phi(\alpha)$ is the classical {\it phase function} ($= E(\alpha)/E(0)$), $R_p$
is the planet's radius, and $a$ is its orbital distance.  The planet/star flux
ratio is the central formula of this paper, and the calculation of $A_g$ and
$\Phi(\alpha)$ is our major focus.

The {\it spherical albedo} is the fraction of incident light reflected by a sphere
at all angles.  For a theoretical object with absolutely no absorptive opacity, all incident
radiation is scattered, resulting in a spherical albedo of unity.  The
spherical albedo is related to the geometric albedo by
$A_s=qA_g$, where $q$ is the {\it phase integral}:
\begin{equation}
q = 2\int_0^{\pi}\Phi(\alpha)\sin\alpha d\alpha.
\label{eq_psratio}
\end{equation}
For isotropic surface reflection (Lambert reflection) $q = \frac 3 2$, while
for pure Rayleigh scattering $q = \frac 4 3$.
Although not written explicitly, all of the above quantities
are functions of frequency ($\nu$).

\subsection{Keplerian Elements}
\begin{figure}
\plotone{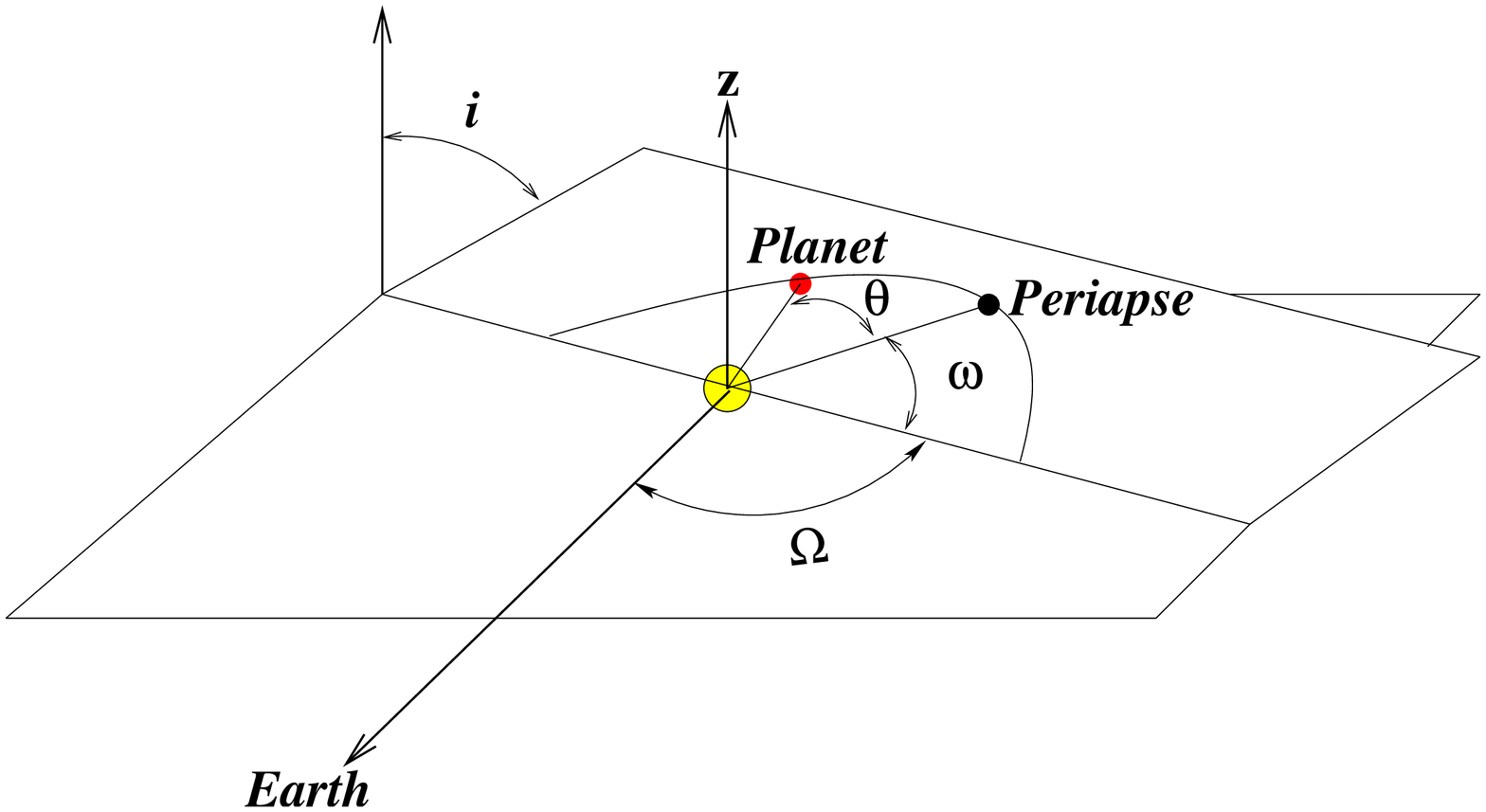}
\caption{Geometry of a planet in orbit about its central star.  The
Keplerian elements include the inclination ($i$), argument of
periastron or periapse ($\omega$), longitude of the ascending node
($\Omega$), and orbital angle or ``true anomaly'' ($\theta$). See text
for details.
\label{fig_diagram}}
\end{figure}

In order to produce a model light curve for a planet orbiting its central star,
one must relate the planet's orbital angle ($\theta$),
as measured from periastron (periapse), to the time ($t$) in the planet's
orbit (Fig. \ref{fig_diagram}):
\begin{equation}
\label{eq_time}
t(\theta) = {{-(1-e^2)^{1/2}P}\over{2\pi}}\left({{e\sin\theta}\over
{1+e\cos\theta}} - 2(1-e^2)^{-1/2}\tan^{-1}\left[{{(1-e^2)^{1/2}\tan(\theta/2)}
\over{1+e}}\right]\right),
\end{equation}
where $P$ is the orbital period and $e$ is the eccentricity.
Other important orbital elements include the orbital inclination ($i$), the
longitude of the ascending node ($\Omega$), and the argument of periastron ($\omega$; a.k.a.
argument of periapse).
The orbital inclination ranges from 0$^\circ$ for fully face-on orbits to 90$^\circ$ for
edge-on orbits.  The {\it ascending node} is the point in an orbit of the
south-to-north crossing of the planet through the horizontal plane of the observer's
line of sight.  $\Omega$ is the angle between
the observer's line of sight and the intersection of
the observer's plane and the orbital plane (the line of nodes).  The argument of
periastron is the angular distance measured along the orbit from the ascending
node to periastron.  For $\Omega=90^\circ$, the line
of nodes is perpendicular to the observer's line of sight, and this is
our default value.  For an arbitrary orientation of
an orbit, the phase angle is related to the above orbital elements by
\begin{equation}
\label{eq_elements}
\cos(\alpha) = \sin(\theta+\omega)\sin(i)\sin(\Omega) - \cos(\Omega)\cos(\theta+\omega).
\end{equation}
By combining eq. (\ref{eq_time}) and (\ref{eq_elements}), we derive the exact
phase of any orbit at any time.  

\subsection{Atmosphere Code}
We obtain each EGP atmospheric temperature-pressure (T-P) structure
with our version of the {\it TLUSTY} 1-D atmosphere code (Hubeny \& Lanz 1995),
as described by Sudarsky, Burrows, \& Hubeny (2003).  In this process,
the external radiation is assumed to be isotropic (i.e., it is averaged
over all angles).
For each resulting T-P structure, we employ a new 2-D version of {\it TLUSTY}
(Hubeny 2005) in order to
obtain the reflection coefficient, $\rho(\mu,\mu_0,\phi,\nu)$,
at any latitude and longitude on the planet's ``surface'' (i.e., the atmosphere).
We assume that the planet is spherical and take small areal
patches on the surface that are essentially planar, but that are irradiated
at different angles depending on their latitude and longitude. 
The local patches are assumed to be spatially 1-D (i.e., they are
locally plane-parallel, horizontally-homogeneous layers).  The raditive transfer is
2-D in angle, so that the anisotropic scattering phase function and (strongly)
anisotropic irradiation are treated exactly.
Details of the 2-D radiative transfer technique are given in Appendix \ref{append}. 

Figure \ref{fig_TPprofiles} depicts a
selection of our theoretical T-P profiles, along with the condensation curves
for ammonia and water.  Those model T-P profiles that cross the
condensation curves contain clouds of the respective species.  For example,
in Fig. \ref{fig_TPprofiles}, the T-P profile for the 3 AU model intersects
the H$_2$O condensation curve, but not the NH$_3$ curve.  Therefore, this
EGP contains a water cloud layer, but its ammonia remains in gaseous
form.  For each model, the frequency-integrated flux at the base of the
atmosphere is
set equal to the integrated emergent flux of a 1-\mj, 5-Gyr, non-irradiated EGP,
which has an effective temperature of $\sim$100 K.
(Burrows \etal 1997).  Although such a prescription is
invalid for full evolutionary models of irradiated EGPs, this choice is perfectly
acceptable for the present study; optical/near-infrared phase functions
and light curves are not sensitive to this internal flux because it is low relative to the
incident stellar flux.

With
the resulting wavelength-dependent reflection coefficients, we integrate over the planetary 
surface to obtain the planetary phase function.  With our numerical technique,
we have reproduced the analytic Lambert phase function (Sobolev 1975) and
the Lommel-Seeliger phase function (van de Hulst 1980) to better than 0.1\%.

\subsection{Atmospheric Composition} 
The realistic representation of condensate clouds is important
in the production of accurate planetary model phase functions, because ices
and grains are generally far more reflective and have much sharper
angular phase dependences (single-particle phase functions) than gases.  In this vein, we use a full
Mie theory code (Li \& Greenberg 1997), which computes accurate angular
scattering phase functions up to size parameters of $2\pi a_0/\lambda$ $\sim$6000,
where $a_0$ is the particle radius, and $\lambda$ is the wavelength of light.  Optical
constants are taken from Martonchik \etal (1984; ammonia ice), and Warren
(1984, 1991; H$_2$O ice).

We employ a
cloud prescription that automatically positions the cloud base
at the intersection of the atmospheric T-P profile
and the condensate curve.  The cloud position is updated with each
iteration of the atmosphere code, so that the final converged model
is derived in a self-consistent manner.  A Deirmendjian (1964)
particle size distribution is used, and the actual modal particle size is
determined numerically using the prescription of Cooper et al. (2003).  For
simplicity, the vertical extent of each cloud layer is set to one pressure
scale height, an approximation that is guided by the cloud model prescription
used.  Adaptive numerical zoning is used to resolve the cloud carefully from top to
bottom.  Previous investigations selected particle sizes more arbitrarily and,
in some cases, distributed these particles homogeneously throughout
the atmosphere (i.e., created clouds with infinite scale heights).

Gaseous opacities include those from the set of atomic and molecular
species described in Burrows \etal (2001) and Sudarsky, Burrows, \& Hubeny (2003).
We use the rainout prescription of Burrows \& Sharp (1999) to account for
condensation and settling in a gravitational field.  Elemental solar abundances
are assumed.

\subsection{Reflection off an EGP Atmosphere}
\label{sec_3D}
\begin{figure}
\plotone{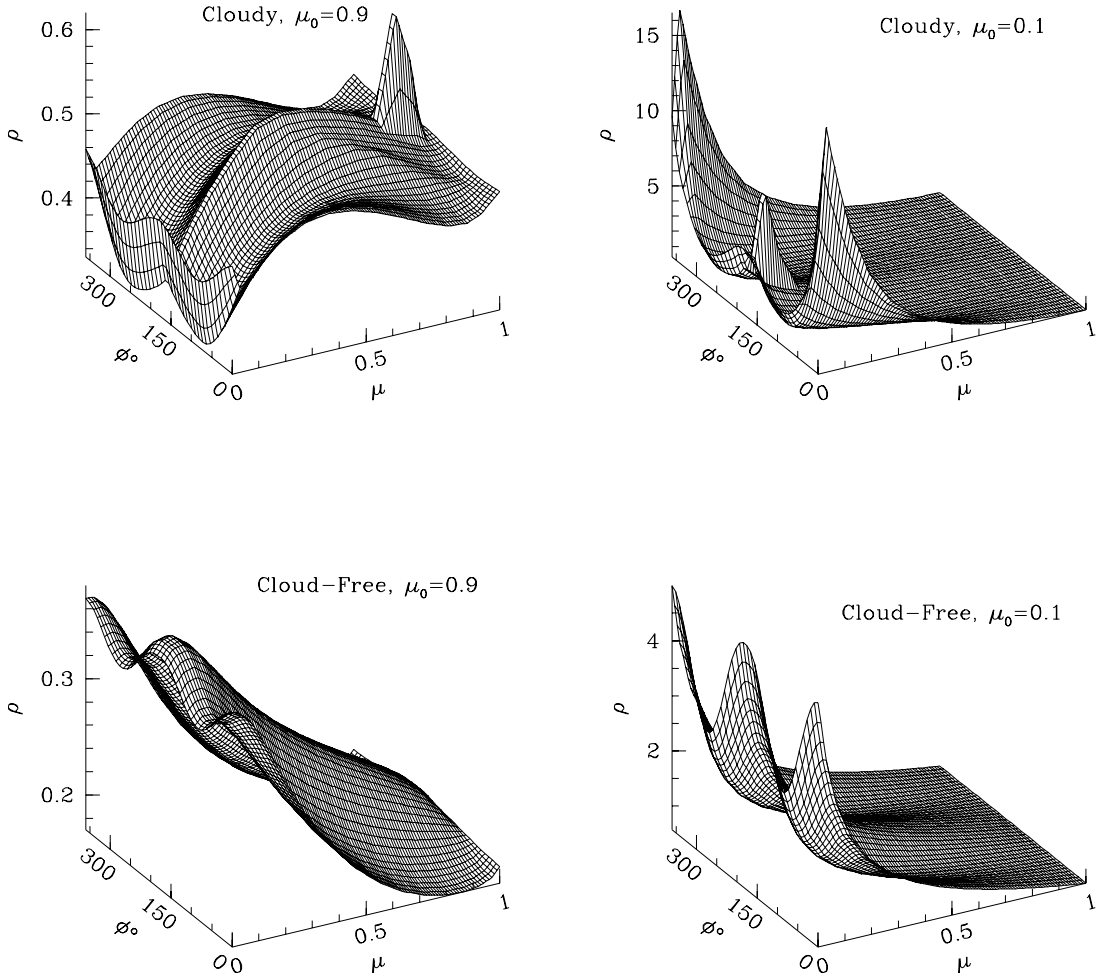}
\caption{The scattering of incident 0.55 $\mu$m radiation off cloudy (condensed ammonia) and cloud-free
atmospheres.  For a given angle of incidence (the cosine of which is represented by $\mu_0$) the reflection
coefficient ($\rho$) is plotted versus the cosine of the emergent angle ($\mu$) and the
azimuthal angle ($\phi$).
{\bf a: (upper left panel)} For a moderate incident
angle ($\mu_0$=0.9, which is $\sim$26$^{\circ}$ from the normal to the surface)
onto a cloudy atmosphere, a backscatter peak is evident, but there is only a
modest variation overall in the reflection coefficient with $\mu$ and $\phi$. 
{\bf b: (upper right panel)} For oblique angles of incidence, radiation
emerges in a strong oblique manner at all azimuth values, but particularly
in the forward direction. {\bf c: (lower left panel)}  For the $\mu_0$ = 0.9
cloud-free case, the reflection coefficient again
does not vary enormously with $\mu$ and $\phi$, although it is clear that more
radiation emerges at oblique angles than along the normal. {\bf d: (lower right panel)}
In the $\mu_0$ = 0.1 cloud-free case,
radiation scatters very obliquely, but unlike the cloudy case, the strength of the scattering in
the forward and backward directions is nearly equivalent.
\label{fig_sm3d4}}
\end{figure}

Since the nature of the reflection of stellar light from an EGP atmosphere
determines the planetary phase function, we explore the reflection properties
in some detail.  Reflected light from giant planets is due to Rayleigh scattering
by gases and/or scattering by the condensates in a planet's atmosphere.
We illustrate the dependence of the optical (0.55 $\mu$m) reflection coefficient
on angle of incidence (which corresponds to planetary latitude and longitude
according to eq. (\ref{eq_latlong})) and the presence or absence of clouds
in the upper atmosphere.  Figures \ref{fig_sm3d4}a through \ref{fig_sm3d4}d
show the reflection coefficient ($\rho$) versus the cosine of the emergent angle
($\mu$) with respect to the normal and the azimuthal angle ($\phi$) for a given incident
angle (the cosine of which is represented by $\mu_0$).
Figure \ref{fig_sm3d4}a depicts $\rho$ for a moderate incident
angle ($\mu_0$=0.9, which is $\sim$26$^{\circ}$ from the normal to the surface) for
an atmosphere with an ammonia cloud layer (modal particle size of
50 $\mu$m).  For such an incident angle, a backscatter peak is evident, but
there is only a modest variation overall with $\mu$ and $\phi$
in the reflection coefficient.  This result contrasts
sharply with that for an oblique angle of incidence ($\mu_0$ = 0.1;
Fig. \ref{fig_sm3d4}b; note the vertical scale change), where the emergence
at oblique angles is much stronger than along the normal to the surface.
Also, reflection off the surface in the forward direction ($\phi$ = 0$^{\circ}$) is
significantly stronger than in the backward direction ($\phi$ = 180$^{\circ}$).  How do these scattering
results compare with those for a cloud-free atmosphere?  For the $\mu_0$ = 0.9
cloud-free case (Fig. \ref{fig_sm3d4}c), the reflection coefficient again
does not vary enormously with $\mu$ and $\phi$, although it is clear that more
radiation emerges at oblique angles than along the normal.  In the $\mu_0$ = 0.1 cloud-free case,
radiation scatters very obliquely, but unlike the cloudy case, scattering in
the forward and backward directions off the surface is nearly equivalent.

In general, for radiation incident at angles relatively close to the
normal, the scattering will not vary greatly as a function of $\mu$ or
$\phi$, but it varies enough that Lambert (i.e., isotropic) reflection is a poor approximation.
For more oblique angles of incidence, forward and backward
scattering off the surface dominates, with very little reflection into angles
near the normal to the surface.  For such oblique angles of incidence, the forward
and backward emergence off the surface is nearly equivalent in the cloud-free
case, but the forward scattering is significantly stronger in the cloudy case.
Of course, this is all a function of wavelength as well, and so the problem is somewhat
more complex than we can present in this subsection.  Various gaseous
and condensed species will affect the magnitude of the reflection coefficient, depending
upon cloud scattering albedos and gaseous absorption bands.

\section{Solar System Phase Functions}
\label{sec_solarphase}
The classical planetary phase function indicates the fraction of light, relative to full
phase, received by a distant observer from a planet as a function of its phase
angle.  Measurements of phase functions for objects in our Solar System
precede even the earliest space missions.  However,
sufficient phase data could not be obtained for the outer planets before
such missions, due to the fundamental limitations of our vantage point
from Earth.  

\begin{figure}
\plotone{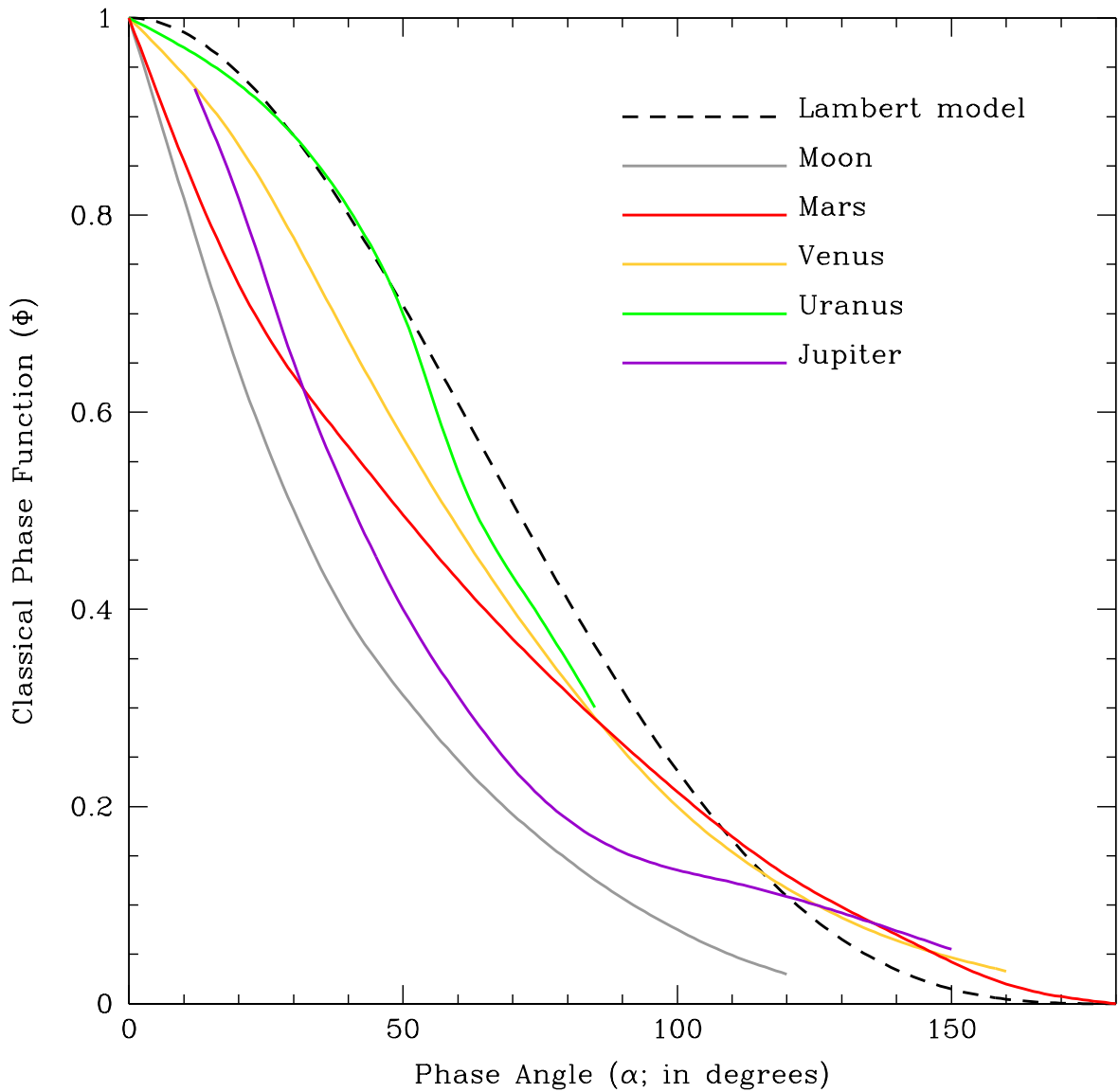} 
\caption{The measured visual phase functions for a selection of Solar System objects.
A Lambert scattering phase curve,
for which radiation is scattered isotropically off the surface regardless of
its angle of incidence, is shown for comparison.  The phase functions of the Moon
and Mars peak near full phase (the so-called ``opposition effect'').  A red bandpass Jupiter
phase function, taken from Dyudina \etal (2004), is also plotted.
\label{fig_solarphase}}
\end{figure}

Figure \ref{fig_solarphase} depicts optical phase functions, $\Phi(\alpha)$,
for a selection of Solar System objects along with that of a Lambert model.
By definition, the classical phase
function is normalized to unity at full phase.  Hence, no albedo information
is given in this figure, but it is useful in understanding the nature
of the scattering itself.  For solid bodies with thin atmospheres, such
as Mars ({\it red curve}; Thorpe 1976), or no atmosphere (e.g. the Moon;
{\it gray curve}; Lane \& Irvine 1973), backscattering can be significant.  Near
full phase the backscattering
contribution is greatest, and this ``opposition effect''
is seen in these phase functions, which peak rather strongly near full phase.
In contrast, consider the Lambert
scattering case ({\it black dashed curve}),
for which radiation is scattered isotropically off a surface regardless of
its angle of incidence.  In that case, the phase function is more rounded
near full phase.  Lambert scattering
appears to be a fair approximation for some objects, such as Uranus
({\it green curve}; Pollack \etal 1986), although such isotropic reflection is
not manifested by any real object.   Note that isotropic reflection off
a surface should not be confused with an isotropic single-particle {\it scattering} phase
function.  The latter generally does not result in isotropic reflection.

Despite the number of missions to Jupiter and Saturn, full planetary phase
data were never made available.  However, Dyudina \etal (2004) recently 
constructed planetary phase functions for these planets by fitting a two-term Henyey-Greenstein
function to original {\it Pioneer} red bandpass scattering data taken at several phase
angles.  The only ambiguities in their resulting planetary phase functions
are at phase angles greater than $\sim$150$^\circ$ (crescent phase) or
less than $\sim$10$^\circ$ (near full phase), because no {\it Pioneer} data were available from these
regions.  Their red-bandpass phase function for Jupiter is plotted in Fig. \ref{fig_solarphase}.

\section{EGP Phase Functions}
\label{sec_EGPphase}
\begin{figure}
\plotone{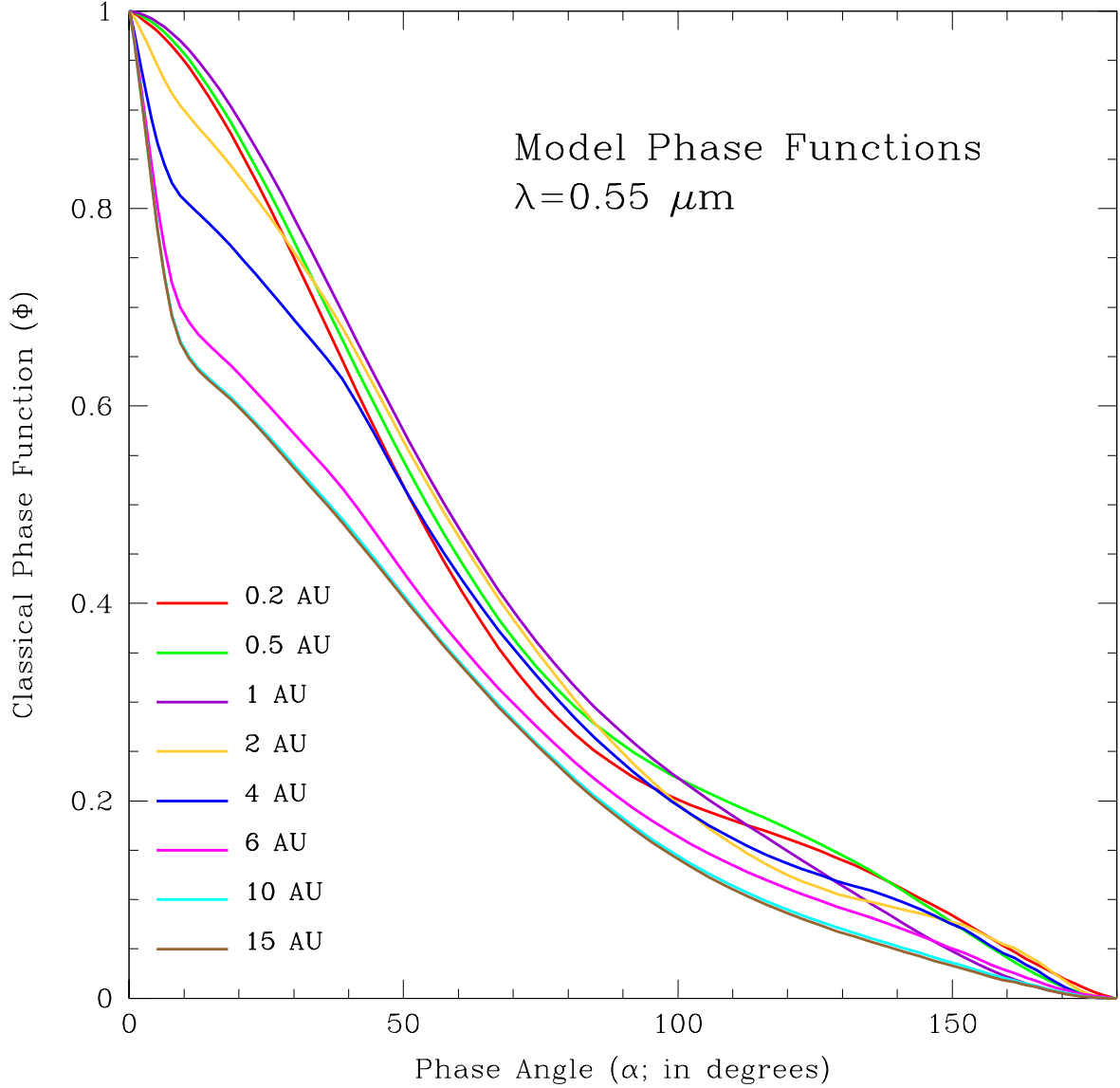} 
\caption{Theoretical optical phase functions of 1-\mj, 5-Gyr
EGPs ranging in orbital distance from 0.2 AU to 15 AU from a G2V star.  Near full phase, the
phase functions for our baseline models at larger orbital distances peak most strongly.
For the cloud-free EGPs at smaller
orbital distances (0.2 AU, 0.5 AU, and 1 AU), the phase functions are more rounded near
full phase.
\label{fig_phaseall}}
\end{figure}

EGP phase functions are determined essentially by the constituents of
a planet's atmosphere.  Rayleigh scattering dominates
purely gaseous atmospheres, while grains and ices often result in strong
forward and backward scattering peaks.
Our theoretical EGP optical ($\lambda$ = 0.55 $\mu$m) phase functions are shown in Fig. \ref{fig_phaseall}.
Included are 1-\mj, 5 Gyr planets ranging in orbital distance from 0.2 AU to 15 AU about a G2V central
star.  Due to a low atmospheric temperature, the model planets
beyond $\sim$4.5 AU contain an ammonia cloud layer above a deeper water cloud deck (i.e., class I;
Sudarsky, Burrows, \& Pinto 2000; Sudarsky, Burrows, \& Hubeny 2003; Burrows,
Sudarsky, \& Hubeny 2004).  Unlike Jupiter and Saturn,
virtually all currently known EGPs, including the long-period Epsilon Eridani and 55 Cancri
planets, likely are too warm to contain condensed ammonia (Sudarsky, Burrows, \& Hubeny 2003;
Burrows, Sudarsky, Hubeny 2004).
However, these wide-separation EGPs will contain condensed H$_2$O (i.e. class II),
as do our models shown at 2 AU and 4 AU.  The theoretical phase functions
for our baseline cloudy EGP models peak (to varying degrees) near full phase
in the optical, the so-called ``opposition effect.''

\begin{figure}
\plotone{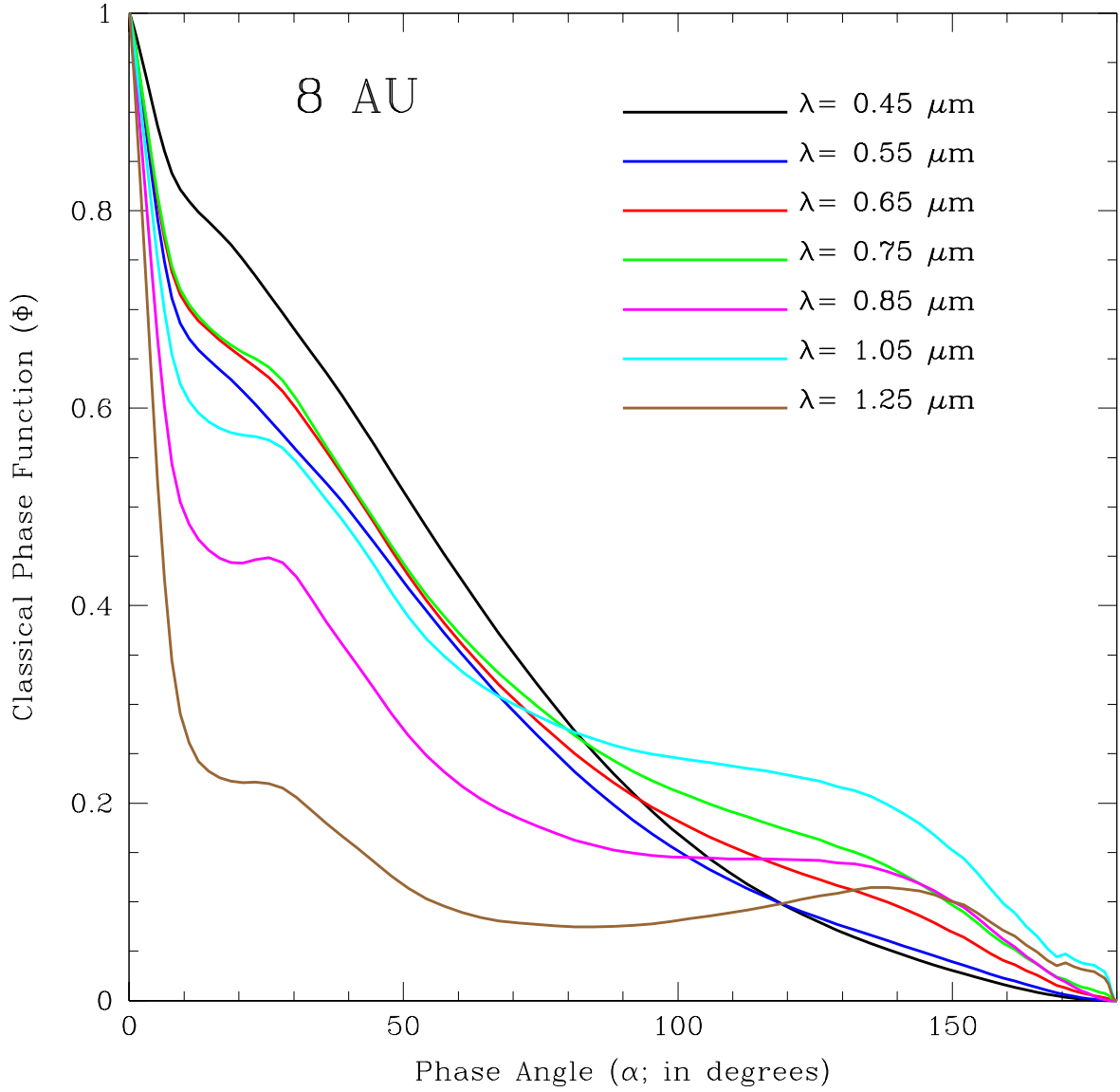} 
\caption{Wavelength dependence of the phase function for an EGP orbiting at a
distance of 8 AU from its G2V central star.  The EGP contains an ammonia cloud
layer above a deeper water cloud.
\label{fig_phaselam}}
\end{figure}

\begin{figure}
\plotone{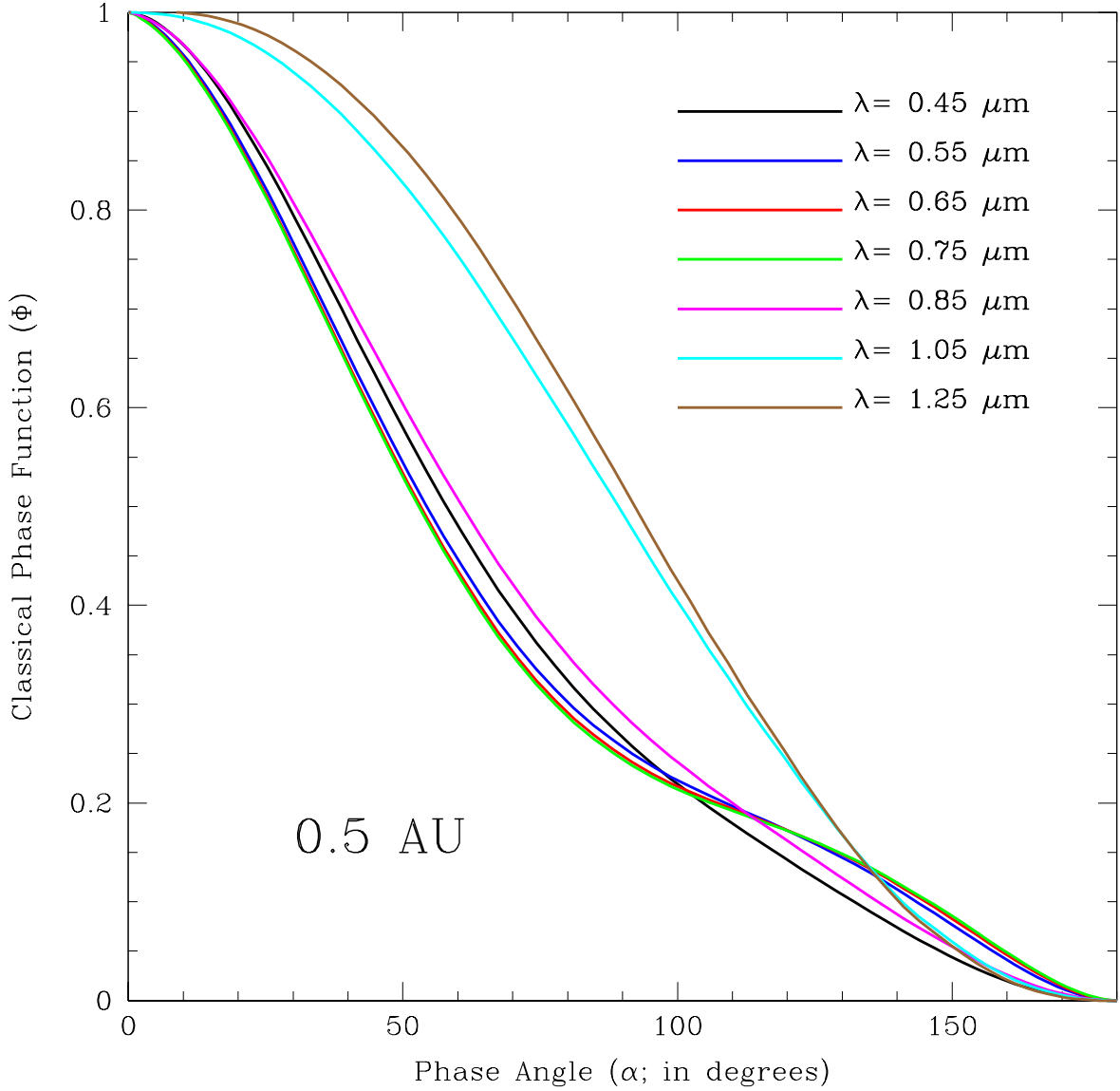} 
\caption{Wavelength dependence of the phase function for a cloud-free EGP orbiting at a
distance of 0.5 AU from its G2V central star.  The 1.05 $\mu$m and 1.25 $\mu$m phase
curves are outliers because they contain a mix of thermally re-emitted and reflected
radiation.
\label{fig_phaselam0.5}}
\end{figure}

An EGP phase function is a wavelength-dependent quantity.  Figure \ref{fig_phaselam} shows
this dependence between 0.45 $\mu$m and 1.25 $\mu$m for our model EGP at 8 AU.  
The general shape of the
phase function is a function of the depth dependence of the scattering and absorption
opacities, which
is a complex function of wavelength.  The wavelength dependence for a
cloud-free model at 0.5 AU is shown in Fig. \ref{fig_phaselam0.5}.  Note that
the 1.05 $\mu$m and 1.25 $\mu$m phase
curves are outliers because they contain a mix of thermally re-emitted and reflected
radiation.

\section{Albedos}
\label{sec_albedos}
\begin{figure}
\plotone{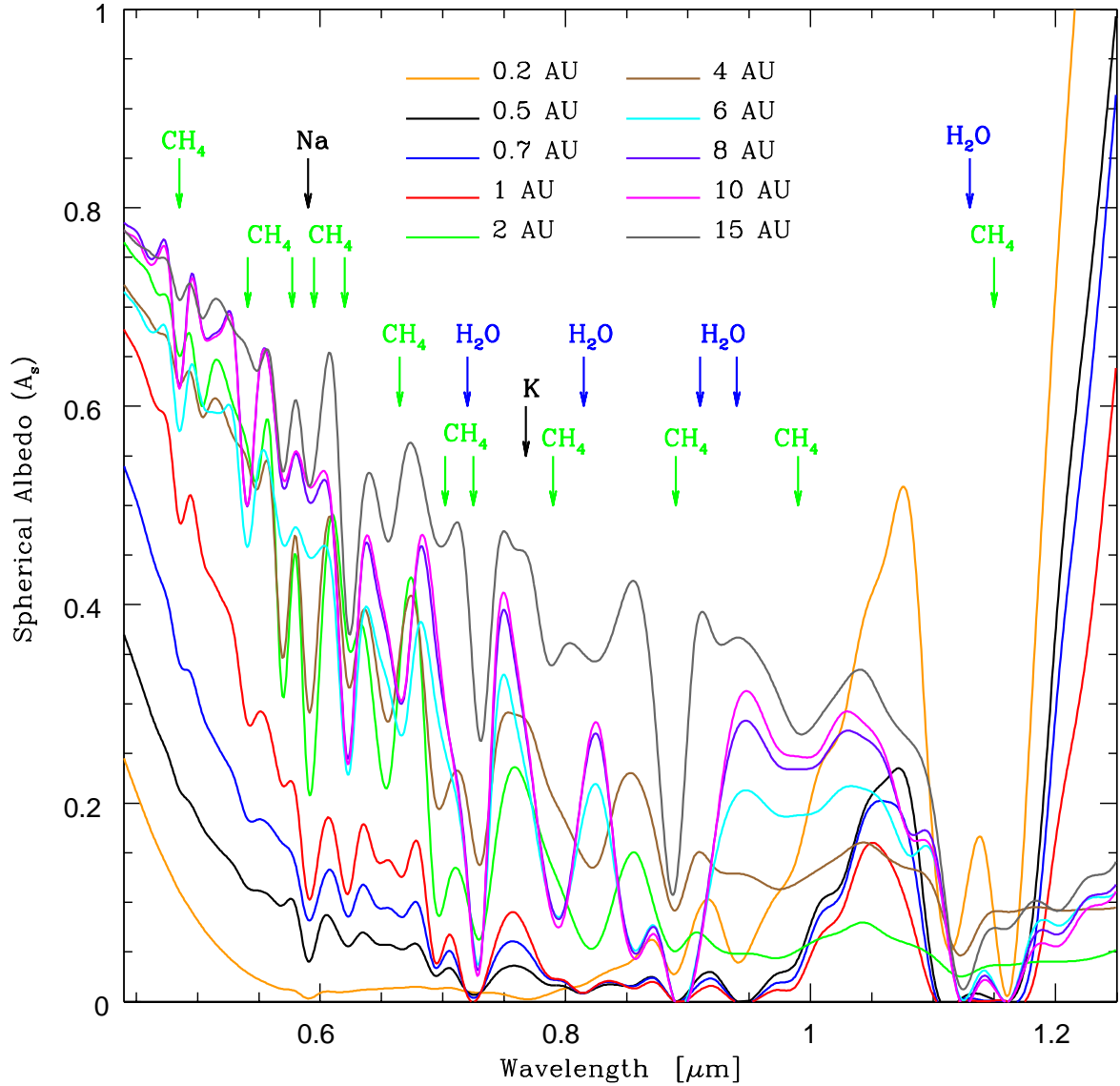} 
\caption{Low-resolution, wavelength-dependent spherical
albedos of 1-\mj, 5 Gyr EGPs ranging in orbital distance from 0.2 AU to 15 AU about
a G2V star.  Cubic splines are fit to all albedo data.  The high albedos at
short wavelengths are due to Rayleigh and/or condensate scattering (reddening effects of photochemical
hazes are not incorporated into these models).  At wavelengths longer than
$\sim$1 $\mu$m, the high ``albedo'' values
shown are due mainly to thermal re-emission, not reflection.
\label{fig_sphericalalbedo}}
\end{figure}

The albedos of EGPs vary substantially, both as a function of orbital distance and wavelength.
Figure \ref{fig_sphericalalbedo} depicts low resolution, wavelength-dependent spherical
albedos of 1-\mj, 5-Gyr EGPs ranging in orbital distance from 0.2 AU to 15 AU about
a G2V star.  The planets beyond 1 AU, with upper atmospheric
water or ammonia cloud decks, exhibit the largest optical
albedos.  At smaller orbital radii, the optical albedos decrease, due to
an absence of reflective condensates and the strengthening of atomic sodium and
potassium absorption.  The closest orbit in our model set
is 0.2 AU.  With very deep silicate and iron clouds, this object is essentially
cloud-free, as are the other EGPs out to $\sim$1.5 AU.
The albedos for objects shown with orbital radii of 1 AU or less appear to rise
into the near infrared.  This effect is not due to increased reflectivity at longer wavelengths.
Rather, it is a result of the object's thermal re-emission of absorbed stellar flux.
The absence of high-altitude clouds combined with strong sodium and potassium
opacity keeps the optical albedo low out to a few tenths of an AU.  However,
with increasing orbital radius, the atmospheric temperatures drop and the
alkali metals condense into chlorides (KCl) and sulfides (Na$_2$S), which rain out (Burrows \& Sharp 1999),
giving way to reflective Rayleigh scattering.  Hence, the optical albedo rises
significantly with increasing orbital radius between $\sim$ 0.2 AU and 1 AU
(see Fig. \ref{fig_sphericalalbedo}), even though there are no water clouds
present.  With the onset of water clouds, the optical albedo rises further
still, which is seen clearly in the model at 2 AU.  The onset and thickening
of reflective ammonia clouds at larger orbital distances results in the
highest albedos at most optical and near-infrared wavelengths.

\begin{figure}
\plotone{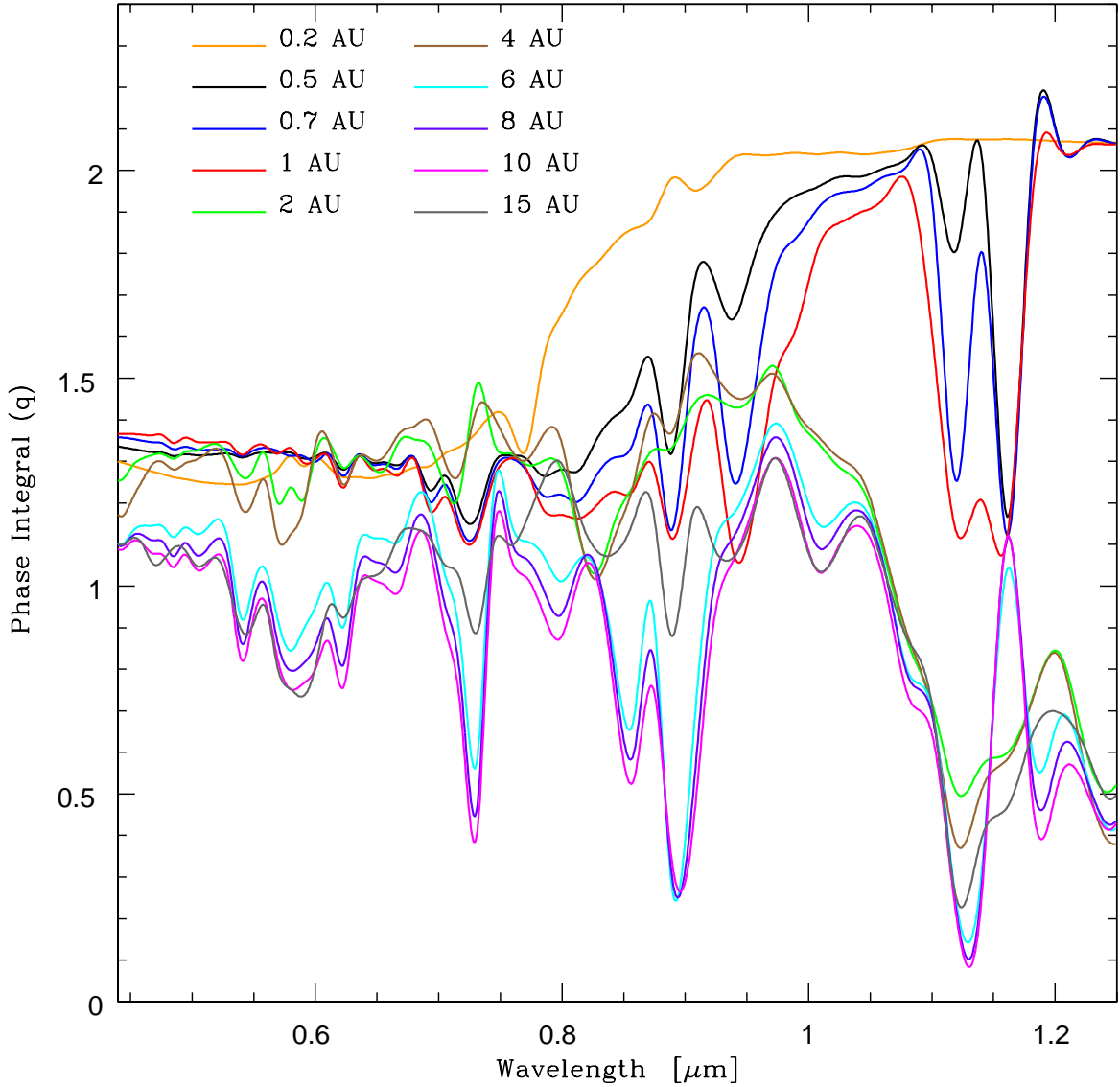} 
\caption{Low-resolution, wavelength-dependent phase integrals of
1-\mj, 5 Gyr EGPs ranging in orbital distance from 0.2 AU to 15 AU about
a G2V star.  Cubic spline curves are fit to the data.  Due to backscattering
effects, cloudy models, particularly
those with ammonia clouds, tend
to have smaller phase integrals.
\label{fig_phaseintegral}}
\end{figure}

\begin{figure}
\plotone{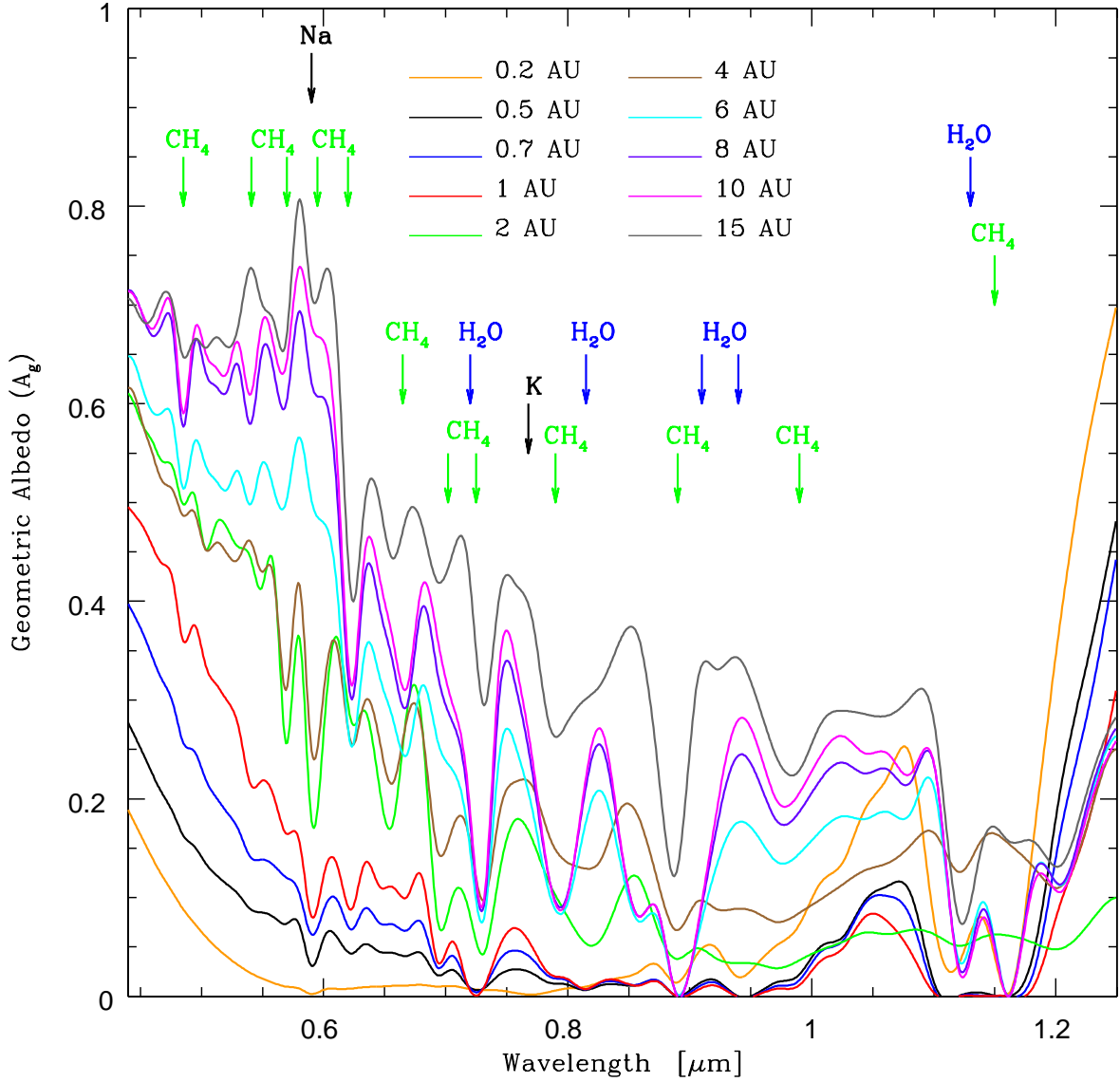}
\caption{Low-resolution, wavelength-dependent geometric
albedos of 1-\mj, 5 Gyr EGPs ranging in orbital distance from 0.2 AU to 15 AU about
a G2V star.  Cubic splines are fit to all albedo data.  As in Fig. \ref{fig_sphericalalbedo},
reddening effects of photochemical
hazes are not incorporated.
\label{fig_geometricalbedo}}
\end{figure}

The geometric albedo is obtained by taking the quotient of the spherical
albedo and the wavelength-dependent phase integral.  Figure
\ref{fig_phaseintegral} depicts cubic spline fits to the phase integral ($q$) for each model EGP.
The phase integral varies widely as a function of wavelength and orbital
distance.  Cloudy atmospheres tend to have smaller phase integrals due to
backscattering effects.  Our resulting geometric albedos are shown in Fig.
\ref{fig_geometricalbedo}.  For comparison,
perfect Lambert reflection (isotropic reflection with no absorption) off a
sphere results in a geometric albedo
of $\frac 2 3$, while it is $\frac 3 4$ for pure Rayleigh scattering.

\section{Light Curves for Circular Orbits}
\label{sec_lightcurves}
Throughout its orbit, an EGP will exhibit different phases with
respect to the observer.  For simplicity, in this section we consider
an idealized case of a circular, highly-inclined orbit of 80$^\circ$.
(An inclination of 90$^\circ$ is edge-on.)
Section \ref{sec_elliptical} investigates the
effects of varying the classical orbital parameters, such as eccentricity,
inclination, argument of periastron, and longitude of the ascending node.

The planet/star flux ratio (eq. \ref{eq_psratio}) is an important quantity that, along with
angular separation, determines the detectability of EGPs (Trauger \etal 2000, 2001).  Due to reflective
Rayleigh scattering and/or condensate clouds at altitude, the optical
spectral region tends to reflect significantly more starlight than the red/infrared
region.  Here, we do not model non-LTE photochemical effects, which may
produce species that somewhat reduce reflection in the UV/blue portion of the
spectrum (e.g., Jupiter).  Therefore, we shall highlight wavelengths longer than
0.5 $\mu$m.

The planet/star flux ratio in the optical is determined
almost entirely by reflected starlight off an EGP atmosphere.  However, such is
not the case in the infrared, where thermally re-emitted radiation,
combined with some reflection due to clouds, will determine
the ratio.  Hence, although an optical EGP spectrum
is expected to be fully phase-dependent, this may not be the case in the
infrared, where emission may be more isotropic, depending on the efficiency
of advection of heat to the night side of the planet.  For the hotter EGPs ($a \lesssim$1 AU),
the near-infrared is likely to be a combination of reflection and thermally
re-emitted radiation.  These components cannot easily be disentangled in
a consistent manner by observations.
In the near-IR, the accuracy of our theoretical light curves for hot EGPs
may be diminished for phase angles that are far from full phase, because
we do not account for thermal emission from planetary longitudes that are not
illuminated by the central star (the ``night side'' of the planet).

\begin{figure}
\plotone{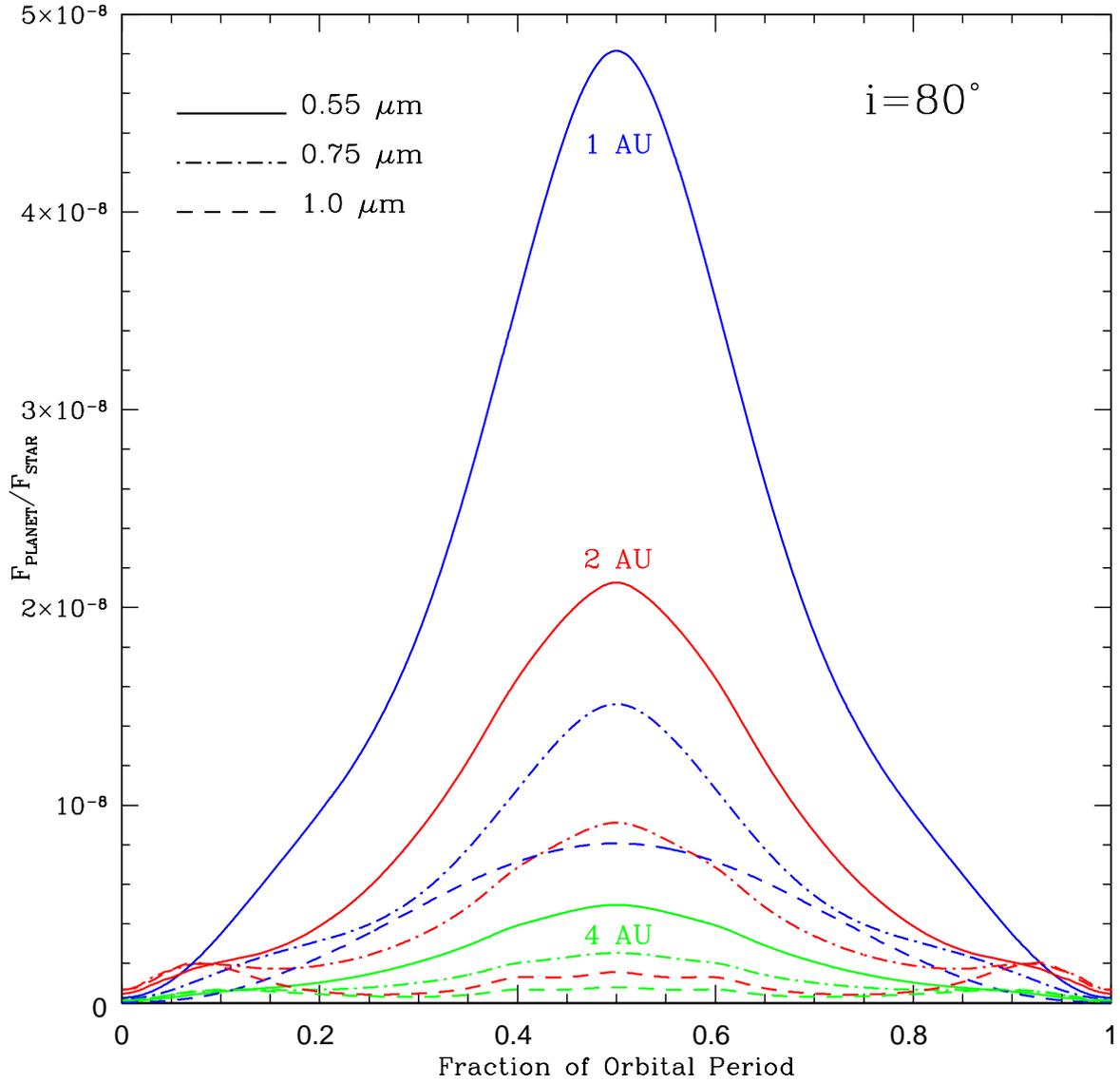} 
\caption{Light curves at 0.55 $\mu$m,
0.75 $\mu$m, and 1 $\mu$m for our model EGPs
in circular orbits inclined to 80$^\circ$ at distances of 1 AU, 2 AU, and 4 AU from a G2V star.  The 
models at 2 AU and 4 AU contain water ice clouds in their upper atmospheres,
while the 1 AU model does not.
\label{fig_lightcurves124}}
\end{figure}

Figure \ref{fig_lightcurves124} shows light curves at 0.55 $\mu$m,
0.75 $\mu$m, and 1 $\mu$m for our model EGPs
in circular orbits at 1 AU, 2 AU, and 4 AU about a G2V star (5 Gyr). The
orbital inclination is set to 80$^\circ$.  The 
models at 2 AU and 4 AU contain water clouds in their upper atmospheres,
while the 1 AU model does not.  The
planet/star flux ratios at full phase range from nearly $5\times 10^{-8}$
at 0.55 $\mu$m for the planet at 1 AU to $\sim$10$^{-9}$ at 1 $\mu$m
for the 4 AU planet.
In addition to their magnitudes, the shapes of the light curves
vary significantly.  For example, the 1 AU model light curve
is broader at 1 $\mu$m than at shorter wavelengths, an effect due
to the mixing of some thermally re-emitted light with the reflected
starlight (as discussed above).  The 2 AU and 4 AU light curve shapes vary
substantially as well, not due to thermal effects, but because
of the strong wavelength dependence of forward scattering off water clouds.
In Fig. \ref{fig_lightcurves124}, the effects of clouds can be discerned by noting
the full-phase flux levels at any wavelength for which there is
no contribution from thermal re-emission.  Comparing
the planet/star flux ratios of the 1 AU
and 2 AU models at 0.55 $\mu$m (or at 0.75 $\mu$m), the brightening of the 2 AU
model due to its water clouds is evident, since applying the inverse square law
to the cloud-free (1 AU) model would produce a significantly lower flux
ratio if it were repositioned at 2 AU.

At 0.55 $\mu$m and 0.75 $\mu$m, the planet/star flux ratio near full phase is
3 to 4 times its value at greatest elongation (its maximum angular separation
as seen from Earth).  In the near-infrared, the variation
from greatest elongation to full phase is much smaller.

\begin{figure}
\plotone{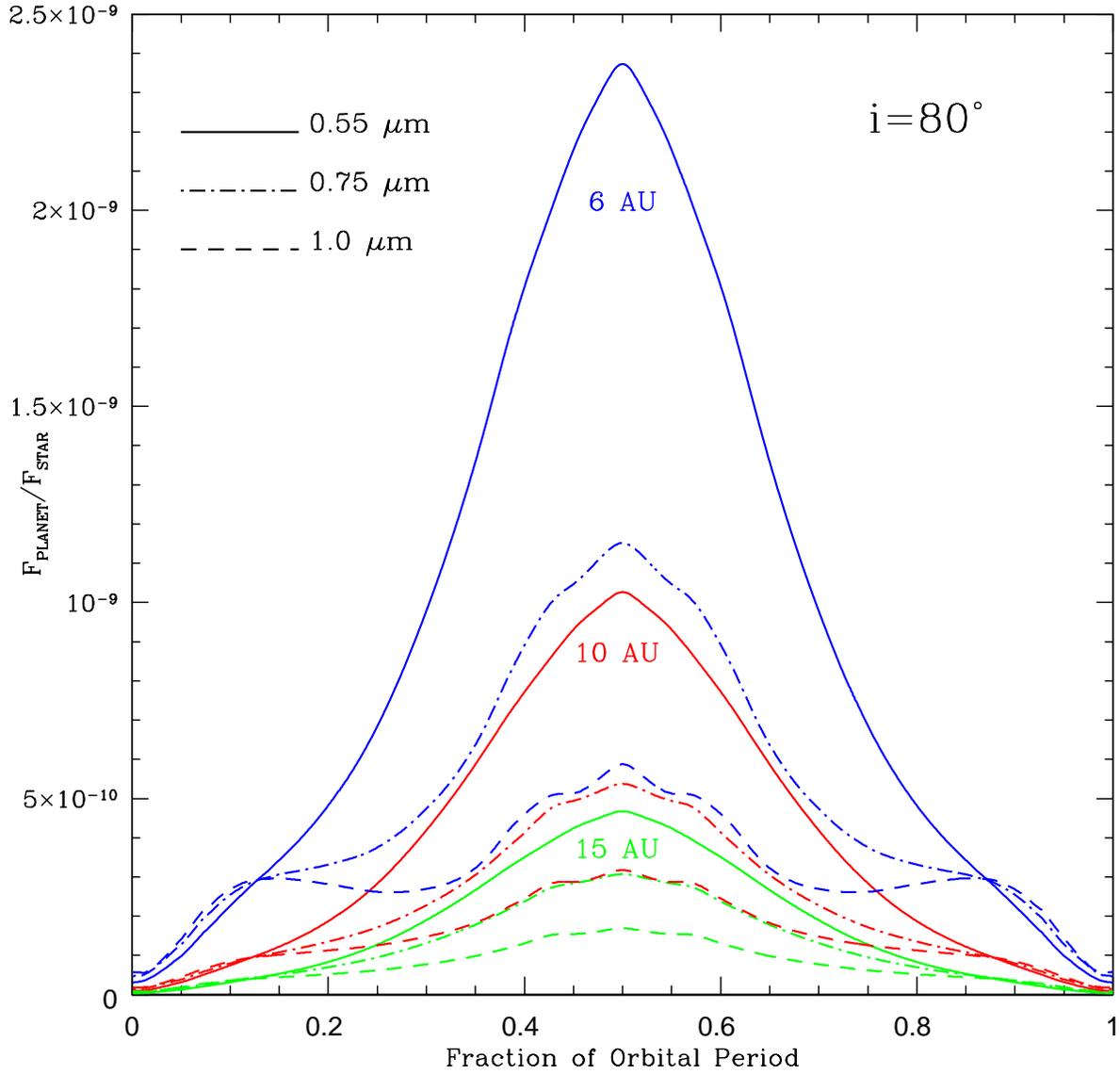} 
\caption{Light curves at 0.55 $\mu$m,
0.75 $\mu$m, and 1 $\mu$m for our model EGPs
in circular orbits inclined to 80$^\circ$ at distances of 6 AU, 10 AU,
and 15 AU from a G2V star.  Each of
these models contains an ammonia ice cloud layer above a deeper water
cloud deck.
\label{fig_lightcurves61015}}
\end{figure}

The light curves for 1-\mj, 5 Gyr ammonia class EGPs (class I) at 6 AU, 10 AU, and 15 AU
about a G2V star are shown in Fig. \ref{fig_lightcurves61015}. 
The largest planet/star flux ratios are in the optical, where an EGP at 6 AU reaches a ratio
of nearly $2.5 \times 10^{-9}$ and our model at 15 AU reaches a value close to
$5\times 10^{-10}$.  At 1 $\mu$m, the peak planet/star
flux ratios are $\sim$20-40\% of their optical values.  As with
the EGPs at 1 AU, 2 AU, and 4 AU, these EGPs with larger orbital radii have planet/star
flux ratios that vary by a factor of 3 to 4 in the optical from greatest elongation to near
full phase.

\begin{figure}
\plotone{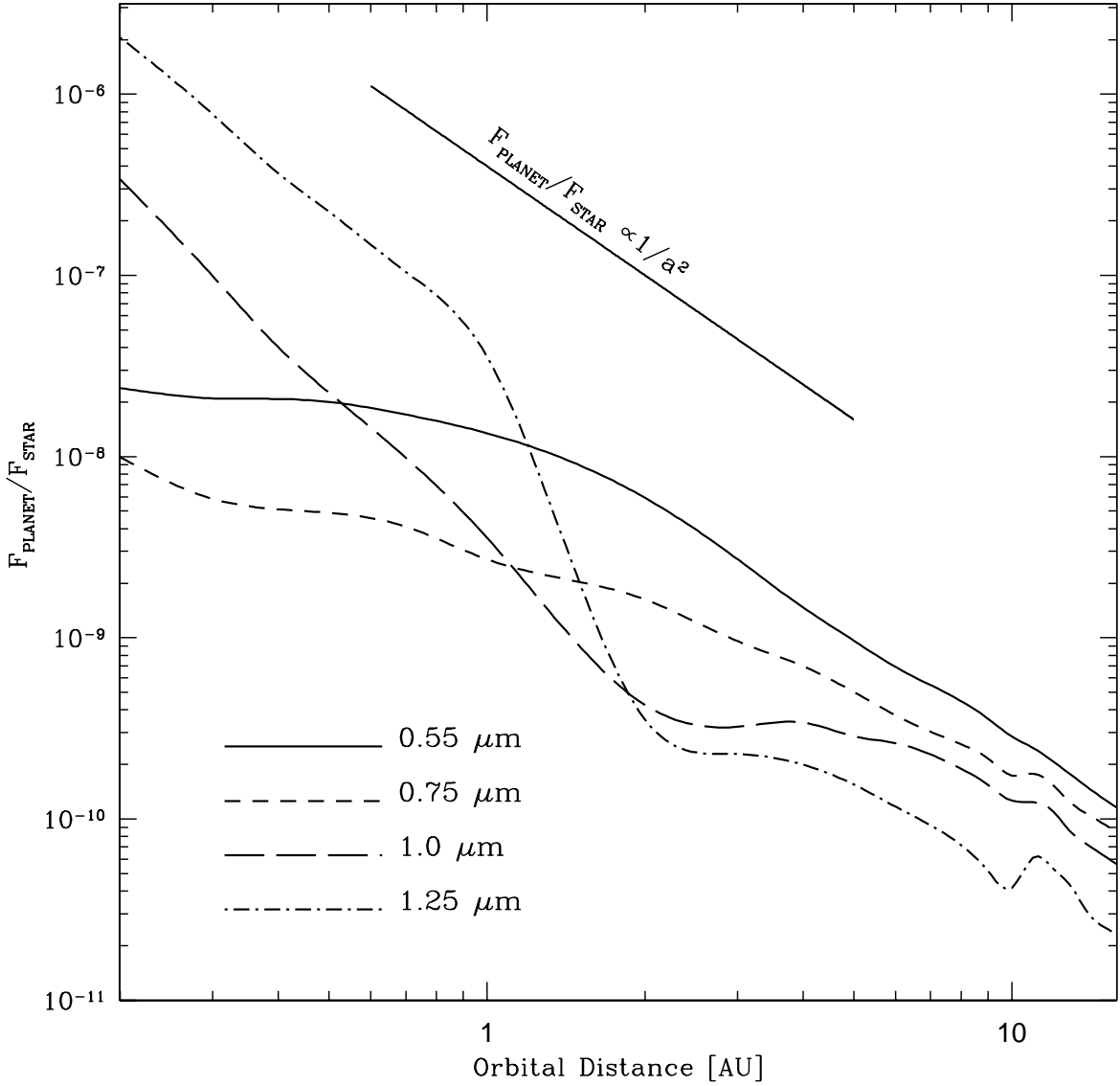}
\caption{Planet/star flux ratio as a function of orbital distance at 0.55 $\mu$m, 0.75 $\mu$m,
1 $\mu$m, and 1.25 $\mu$m assuming a G2V central star.  In each case, the plotted
value corresponds to a planet at greatest elongation with an orbital inclination
of 80$^\circ$.  Note that the planet/star flux ratios do not follow a simple $1/a^2$ law.
\label{fig_ratiodist}}
\end{figure}

The planet/star flux ratio as a function of orbital distance at 0.55 $\mu$m, 0.75 $\mu$m,
1 $\mu$m, and 1.25 $\mu$m assuming a G2V central star is
shown in Fig. \ref{fig_ratiodist}.  In each case, the plotted
value corresponds to a planet at greatest elongation with an orbital inclination
of 80$^\circ$.   For EGPs with relatively
small orbital radii ($\lesssim$ 1 AU), the near-IR flux ratios are large, due to thermal
re-emission of absorbed stellar radiation at these wavelengths.  With increasing orbital
radius, the peak of this thermal re-emission moves to longer and longer wavelengths.  This effect 
coupled with the condensation of reflective clouds beyond $\sim$1.5 AU results in larger
optical flux ratios relative to infrared flux ratios for larger orbital radii.  Note
that the planet/star flux ratios do not follow a simple $1/a^2$ law.

Explicitly averaging our present light curves with respect to phase, we test
the validity of the previous ``phase-averaged'' planet/star flux ratios derived
by Sudarsky, Burrows, \& Hubeny (2003) and by Burrows, Sudarsky, \& Hubeny (2004),
for which a 1-D atmosphere code was used.
Our new phase-averaged results agree quite closely with those
from our previous 1-D treatment, differing by less than 3\% in most cases.

\section{Color Dependence with Planetary Phase}
\label{sec_colors}
\begin{figure}
\plotone{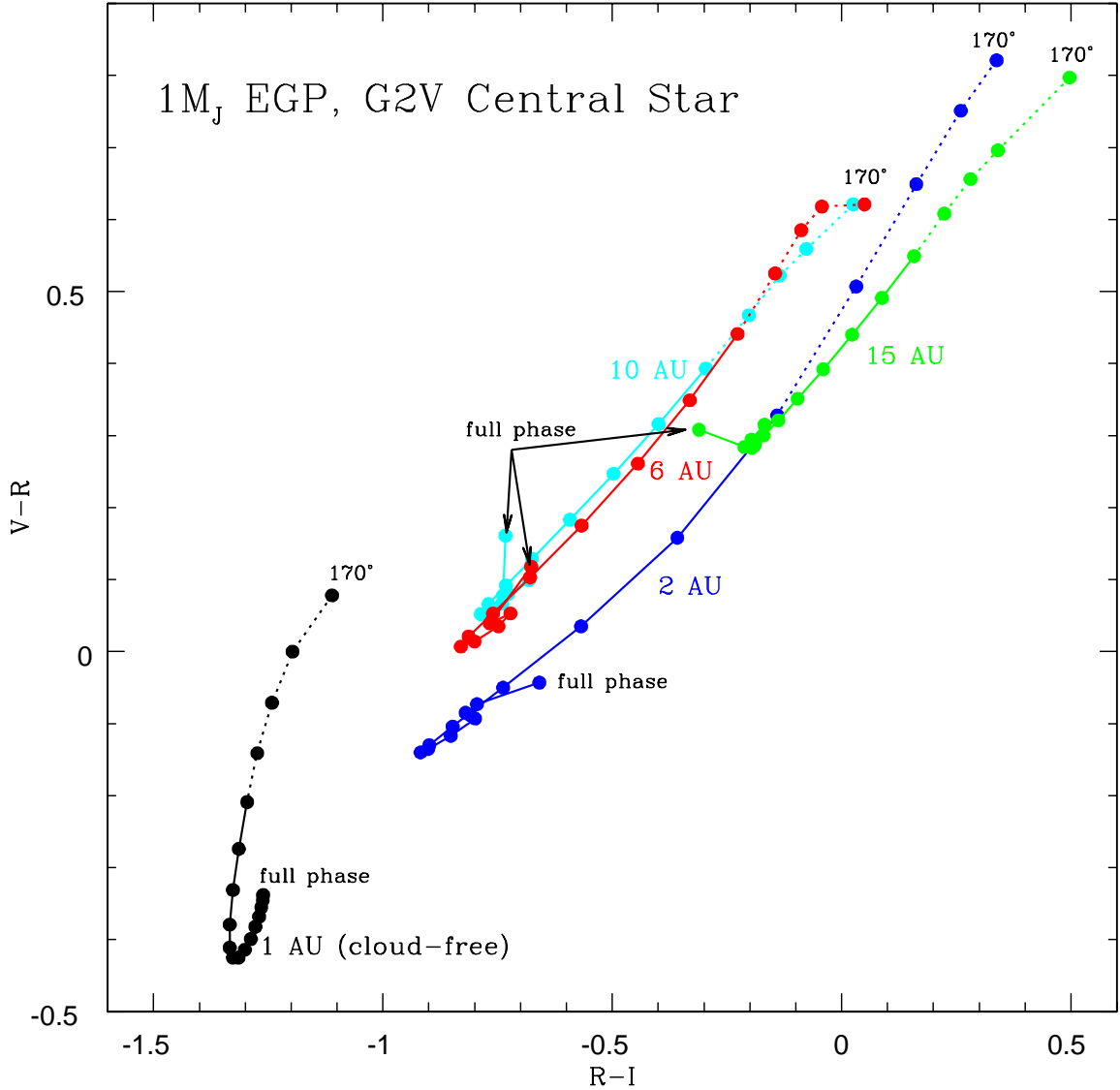}
\caption{$V-R$ vs. $R-I$ color-color diagram detailing variations with
planetary phase for a variety of orbital distances.  Each of the
curves depicts an orbit
from full phase (0$^\circ$) to a thin crescent
phase (170$^\circ$) in increments of 10$^\circ$ (as indicated by the
filled circles).  Cloud-free EGPs are bluest near greatest elongation, while
cloudy EGPs tend to be bluest in a gibbous phase.
As full phase is approached, the colors redden somewhat.  However, the
crescent phases appear to be far redder, varying by as much as
a full astronomical magnitude from their blue gibbous-phase colors in some cases.  See
text for details and a discussion of the accuracy at large phase angles
(denoted by dotted lines).
\label{fig_colorcolor}}
\end{figure}

As shown in \S \ref{fig_phaseall}, EGP phase functions vary with wavelength.
An interesting consequence of this fact is that an EGP is expected
to show color variations throughout the different phases of its
orbit.  Figure \ref{fig_colorcolor} is a $V-R$ vs. $R-I$ color-color diagram,
which details these color variations with planetary phase for a
variety of orbital distances.  Each of the curves in Fig. \ref{fig_colorcolor}
covers an orbit from full phase (0$^\circ$) to a thin crescent
phase (170$^\circ$) in increments of 10$^\circ$ (as indicated by the
filled circles).  For most cloud-free EGPs, the phase that is bluest in both $V-R$
and in $R-I$ is 80$^\circ$ or 90$^\circ$.  That is, cloud-free EGPs are bluest near greatest elongation.
In comparison, EGPs with water clouds and/or ammonia clouds tend to be bluest in a gibbous phase.
As full phase is approached, the colors redden somewhat.  However,
the crescent phases appear to be far redder, varying by as much as
a full astronomical magnitude in some cases.  Bluer
light scatters more efficiently via Rayleigh scattering or cloud
reflection than red/infrared radiation, so an observer viewing an
EGP at an intermediate phase will catch more blue photons than an observer
of a crescent phase, for which the bluer photons have been scattered
away from the line of sight. 

The cloud-free 1 AU model shown in Fig. \ref{fig_colorcolor} deviates significantly 
in color space from the 2 AU (water cloud), 6 AU, 10 AU, and 15 AU models (with ammonia and water
clouds).  The cloudy models are substantially redder in both $V-R$ and $R-I$.  Additionally,
the excursion in $R-I$ is much smaller for the cloud-free model.  This
is not surprising, given that Rayleigh scattering ($\propto\lambda^{-4}$)
is far less effective than condensate scattering in this region of the spectrum.  
Still, in $V-R$, the cloud-free 1 AU model is expected to vary with phase by as much
as 0.5 magnitude.

Due to our use of a 2-D planar code, we
cannot model limb effects, such as transmission of stellar
radiation through a chord of the atmosphere.  Although we are confident
in the color trends of Fig. \ref{fig_colorcolor}, the values at very large phase
angles may not be as robustly calculated as the rest.  We have denoted uncertain
areas of this diagram with dotted lines.  Note that EGP direct imaging
at such large phase angles would be very difficult in these wavelength
bands, due to the lower planet/star flux ratios at crescent
phases and the associated small angular separation from the star.

\section{Condensate Particle Size Dependence}
\label{sec_partsizes}
Modeling cloud particle sizes in EGP atmospheres remains a difficult
endeavor.  The phase functions and light curves presented thus far have been
produced using the prescription of Cooper \etal (2003).  While their
resulting particle sizes are in broad agreement with those of others (Ackerman
\& Marley 2001; Lunine \etal 1989), we recognize that cloud formation in EGPs is a complex
process, which may not be reproduced correctly by current modeling
algorithms.  Furthermore, for simplicity, in this paper we hold the modal particle
size constant throughout the cloud layer, using the particle size at the cloud 
base.  In reality, particle sizes may tend to decrease somewhat toward the
cloud tops, an effect that we have not incorporated.  Hence, we explore the
effects of significant cloud particle size variation on our EGP phase functions.

\begin{figure}
\plotone{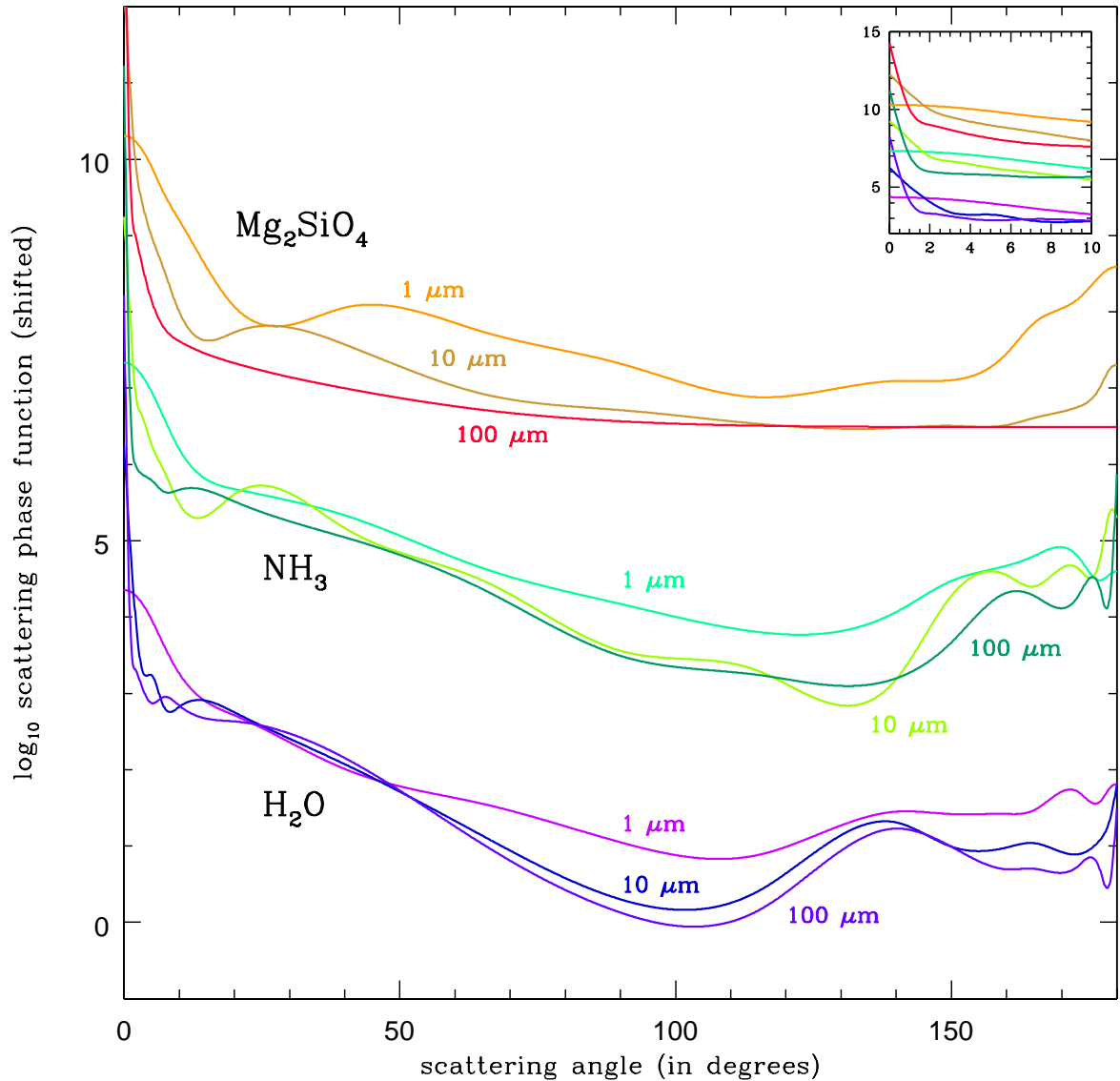} 
\caption{Mie theory optical angular scattering
dependence of H$_2$O ice, NH$_3$ ice, and forsterite grains at modal particle
sizes of 1 $\mu$m, 10 $\mu$m, and 100 $\mu$m on
a logarithmic scale.  A Deirmendjian particle size
distribution is assumed.  Each scattering phase function is normalized to
the 1-$\mu$m particle size (for easy comparison) and fit with a cubic spline.
Furthermore, the various species are offset for readability, so the phase function
values are dimensionless.  The inset figure, also on a logarithmic scale,
shows the degree of forward peaking at small phase angles.
\label{fig_mieang}}
\end{figure}

Figure \ref{fig_mieang} depicts the Mie theory optical angular scattering
dependence of H$_2$O ice, NH$_3$ ice, and forsterite grains at modal particle
sizes of 1 $\mu$m, 10 $\mu$m, and 100 $\mu$m.  Our derived particle sizes are
generally encompassed by this broad range.  Most of our ammonia and (especially) water cloud
modal particle sizes fall near the upper end of the range.
In Fig. \ref{fig_mieang}, the usual Deirmendjian (1964) size distribution function is used.
The data are fit with cubic splines and offset for clarity.
For all species, the angular distribution exhibits a strong
forward peak at small scattering angles.  This peak strengthens with
increasing particle size and is quite extreme at 100 $\mu$m.  Additionally,
a significant backscatter peak is common, an effect that cannot be represented
by the commonly used Henyey-Greenstein scattering phase function.

\begin{figure}
\plotone{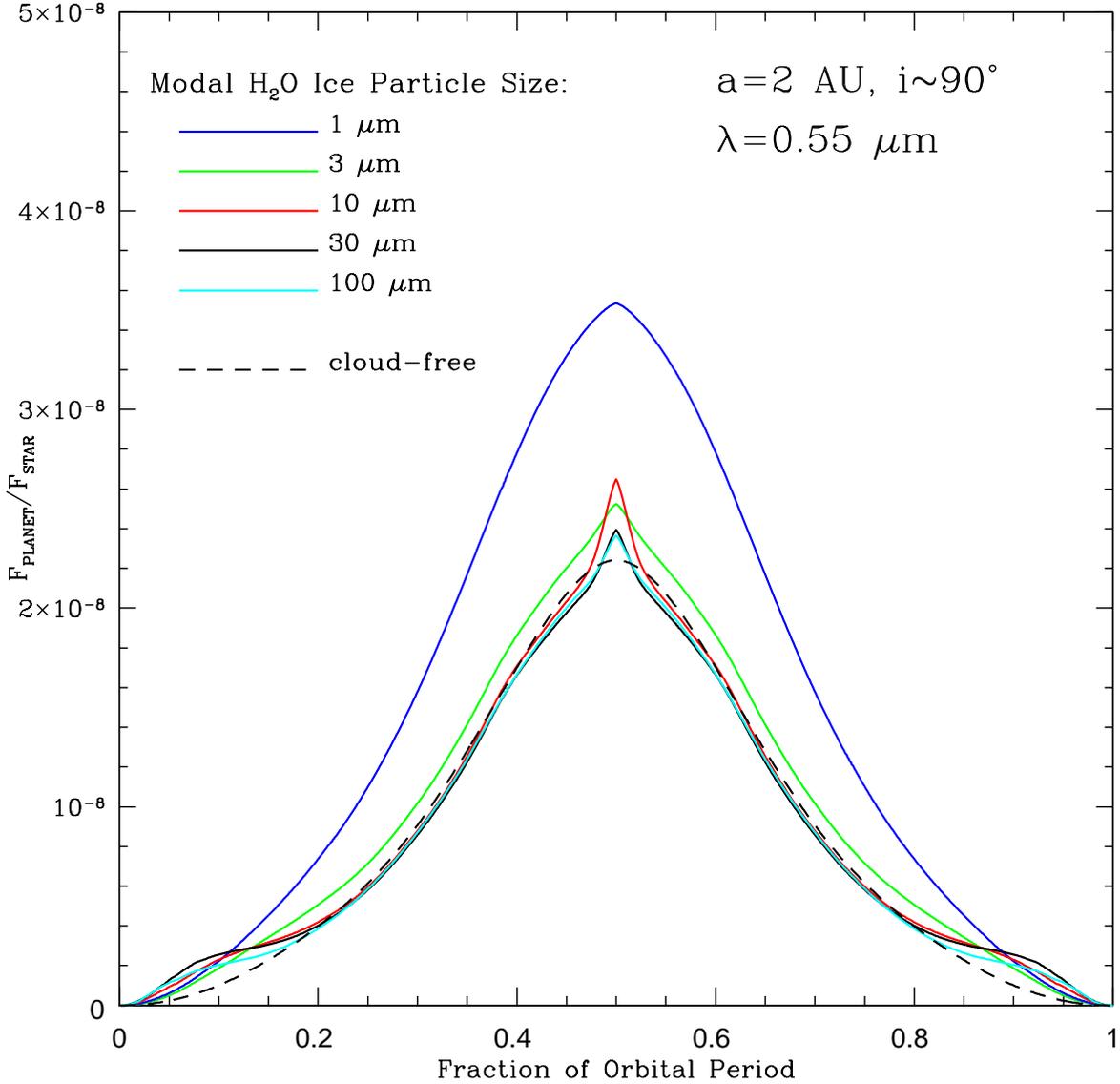}
\caption{The dependence of the planet/star flux ratio on condensate particle
size at a wavelength of 0.55 $\mu$m.  Model light curves for EGPs
at 2 AU with modal H$_2$O ice particle sizes of 1, 3, 10, 30, and 100 $\mu$m are depicted.
Shown for comparison is a cloud-free model ({\it black dashed curve}).
In order to show the full variation in the shapes and magnitudes of the light
curves with particle size, we have set the orbital inclination to
$\sim$90$^\circ$ so that the opposition effect, present for many
of the models, can be seen in full.  Transit effects are not modeled.
\label{fig_sizedep}}
\end{figure}

\begin{figure}
\plotone{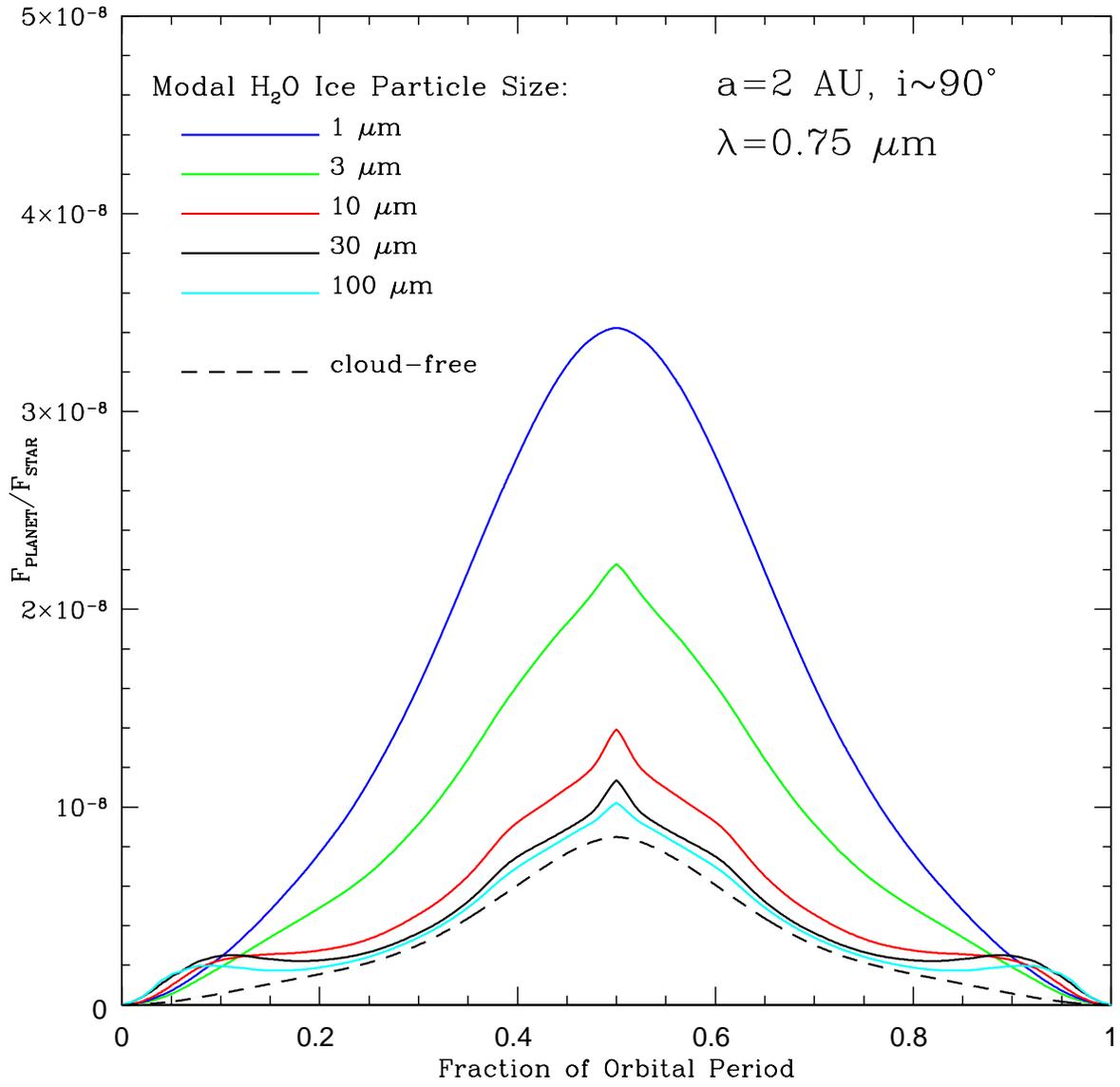}
\caption{Same as Fig. \ref{fig_sizedep}, except at 0.75 $\mu$m.
\label{fig_sizedep2}}
\end{figure}

We explore the dependence of EGP light curves on condensate
particle sizes in Fig. \ref{fig_sizedep} for the
optical wavelength of 0.55 $\mu$m, and in Fig. \ref{fig_sizedep2}
at 0.75 $\mu$m.  In each figure, model light curves for EGPs
at 2 AU with modal H$_2$O ice particle sizes of 1, 3, 10, 30, and 100 $\mu$m are depicted.
Shown for comparison is a cloud-free model ({\it black dashed curve}).
In order to show the full variation with particle size in the shapes and magnitudes of the light
curves, we have set the orbital inclination to
$\sim$90$^\circ$ so that the opposition effect, present for many
of the models, can be seen in full (transit effects are ignored).  Use of Mie scattering theory is
necessary in order to derive the wavelength-dependent scattering properties
for various cloud particle sizes, but smaller particle sizes generally
result in higher planet/star flux ratios at most wavelengths.  Scattering cross sections
are roughly proportional to the square of the particle size, but the
number density of particles for a given condensate mass scales
as the inverse cube of the size.  Hence, the total cloud scattering
opacity tends to be greater for smaller particles at most wavelengths.

At both 0.55 $\mu$m and 0.75 $\mu$m, the 1 $\mu$m particle size models
exhibit higher planet/star flux ratios and smoother light curves than
for those models with larger particles.  At 0.55 $\mu$m, the planet/star flux ratios do not
vary substantially over a range of modal particle sizes from 3 to 100 $\mu$m
(although the shapes of the light curves differ somewhat).  This result contrasts with
those of the same EGPs at 0.75 $\mu$m, where the planet/star flux ratio
becomes progressively lower with increasing particle size.  Such
results indicate the importance not only of condensate particle size, but
of wavelength-dependent light curves as well.

\section{Elliptical Orbits and Dependence on Orientation}
\label{sec_elliptical}
Until now, we have chosen to limit our discussion of EGP phase functions
to nearly edge-on circular orbits.  In reality, EGP orbital inclinations
are randomly distributed, and elliptical
orbits are prevalent for all but the
close-in EGPs.  An observer's viewing angle affects the
range of phases visible, and the eccentricity and semi-major axis determine the
duration of each phase throughout the orbital period.  Strikingly, a
significant eccentricity can lead to a compositional difference
in a planet's atmosphere, because the level of heating from its central
star varies greatly.  We investigate the effects of such changes on the phase
functions and light curves of EGPs.

\subsection{Cloud-Free Elliptical Orbits}
The light curve of a planet in an eccentric orbit will be very different
from that of the same planet in a circular
orbit.  In the cloud-free case, this difference is due largely to
the amount of time the planet spends at a given phase in its orbit
and the varying distance of the planet to its central star at a given
phase.  

Dyudina et al. (2004) pointed out that the peak of the
light curve for an eccentric orbit does not occur at full phase.  This
actually is an effect of the observer viewing angle and is a function of the
argument of periastron, the longitude of the ascending node,
and the orbital eccentricity.  Assuming a default
$\Omega$ of $90^\circ$, the peak does occur
at full phase for $\omega$ = $90^\circ$ or $270^\circ$, but otherwise does not. 

\begin{figure}
\plotone{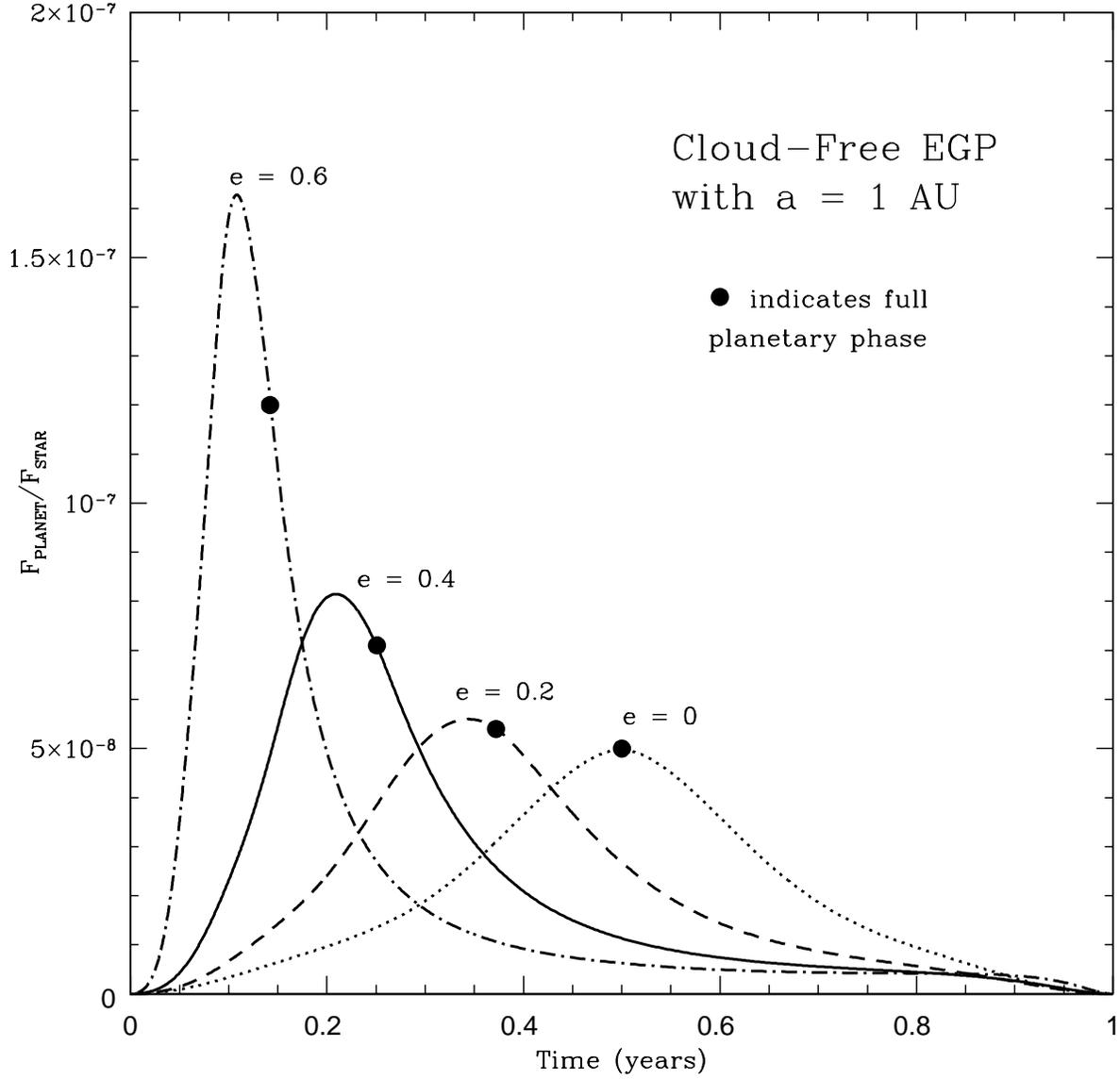}
\caption{The optical planet/star flux ratio as a function of eccentricity
for a cloud-free EGP at 1 AU, fixing $\Omega$ and $i$ at $90^\circ$
and $\omega$ at $0^\circ$ (see text for details).  Only for $e$=0 does the
peak of the light curve coincide with full phase.
\label{fig_eccentricity}}
\end{figure}

Figure \ref{fig_eccentricity} depicts the optical light curves
of orbits with various eccentricities for a cloud-free EGP orbiting a G2V star.
The semi-major axis is set to 1 AU, leading to an orbital period of 1 year.
We use the default values of $\Omega=90^\circ$,
$\omega=0^\circ$ (and $i=90^\circ$ so that full phase can be represented),
resulting in a light curve whose peak
value leads the full phase of the planet. This is because the orbital distance
increases before full phase is reached.  In this figure, the time of
full phase is indicated for each orbit by a filled circle.  Only for a
circular orbit does the peak of the light curve match up with the time
that the planet reaches full phase.  For $e$=0.2, full phase lags the light curve
peak by 13.7 days.  For $e$=0.4, the lag is 18.2 days, while for $e$=0.6 it
is 13.5 days, given our orbital parameter assumptions.  The value of the
light curve maximum varies greatly with orbital eccentricity because
the incident stellar flux on the planet increases by a factor of
$(1+e)^2(1-e)^{-2}$ between apastron and periastron.

\subsection{Condensation and Sublimation Transitions in Elliptical Orbits}
The presence or absence of condensate clouds in the outer atmosphere of
an EGP depends largely on the level of stellar radiation incident on
the planet.  A cloud base resides approximately where the condensation curve of
a given species intersects the atmospheric T-P profile, but the T-P profile
itself is coupled strongly to the intensity of the incident radiation.
Even for eccentricities of 0.2 or 0.3, the
incident fluxes differ enough that an EGP may be virtually cloud-free at periastron,
but have substantial water condensation at apastron.  This ``switching'' of
EGP composition classes will have visible effects in the light curves of such planets.
We model these eccentric orbits by interpolating the albedo and phase spectra
within our grid of circular EGP orbit results between 0.2 and 15 AU.  Hence, we
assume that the timescale of the chemistry associated with cloud formation
and sublimation is much shorter than a planetary orbital period.

\begin{figure}
\plotone{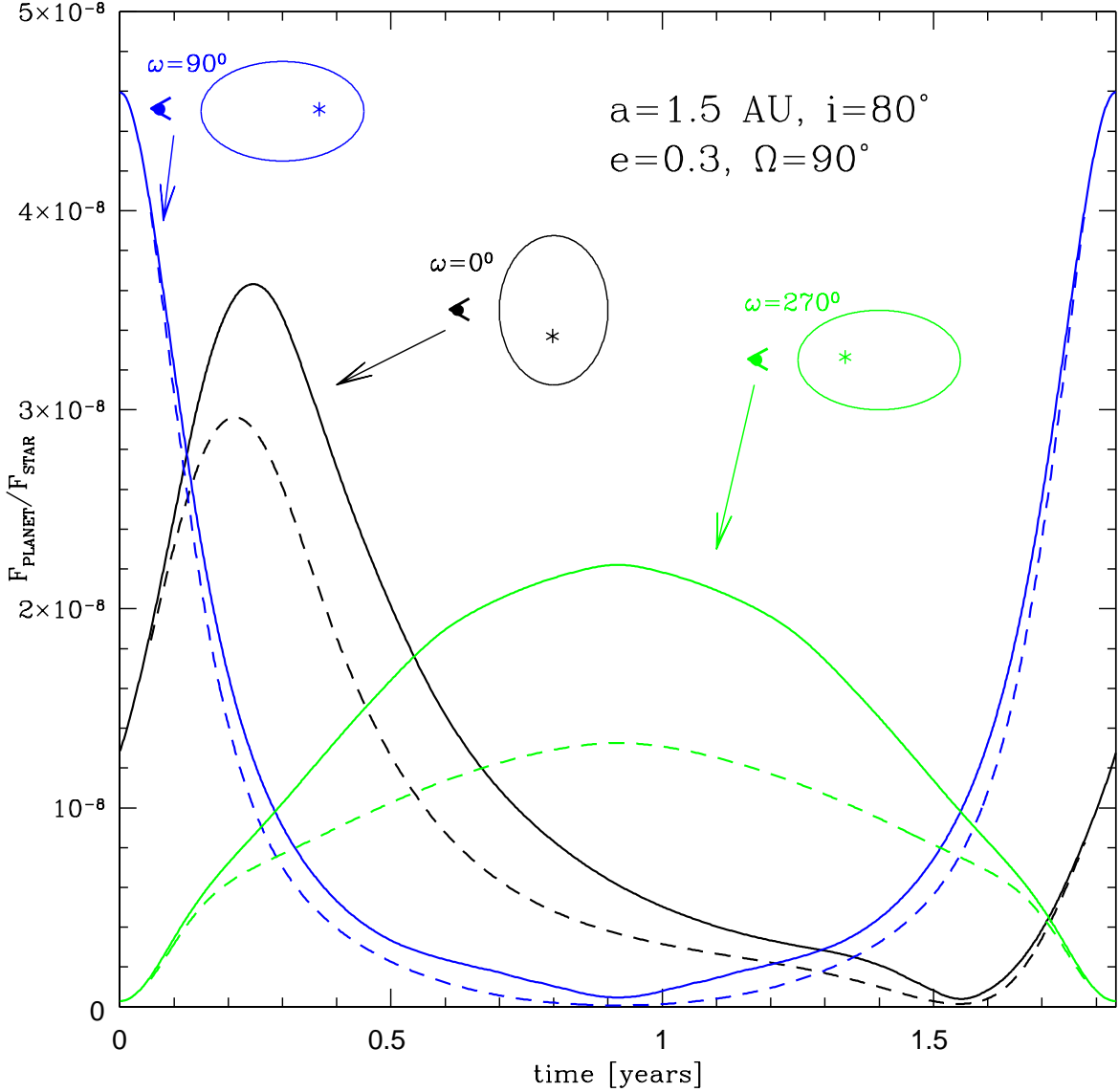} 
\caption{The optical light curve of an EGP with $a$=1.5 AU
and $e$=0.3, assuming a G2V central star (a system similar to HD 160691; Jones
\etal 2002) and an inclination of 80$^\circ$ (with no transit effects).
Three different possible viewing
angles are shown ({\it solid curves}).  Clouds
condense and sublimate throughout each orbit as the planet-star distance
varies in time.  Shown
for the sake of comparison are the light curves that would result if the object
were to remain artificially cloud-free throughout its entire orbit ({\it dashed curves}).
\label{fig_viewingangle}}
\end{figure}

Figure \ref{fig_viewingangle} depicts optical light curves for an EGP with $a$=1.5 AU
and $e$=0.3, assuming a G2V central star (a system similar to HD 160691; Jones
\etal 2002).  Three different possible viewing
angles are shown ({\it solid curves}).  For the sake of comparison, also
shown are the light curves that would result if the object
were to remain cloud-free throughout its entire orbit ({\it dashed curves}), which
is not expected to be the case in reality.  To keep things simple, we have fixed
$i$ at 80$^\circ$ and $\Omega$ at $90^\circ$.  For the three values of $\omega$ shown, both the $0^\circ$
and $270^\circ$ viewing angles exhibit pronounced effects due to
the condensation of water.  At $\omega = 270^\circ$,
the full phase difference between our
actual model with water condensation and one that artificially remains cloud-free
is nearly a factor of 2 in the planet/star flux ratio.  In general, the peak level of
this ratio also depends on the orbital distance of the planet near full phase.
In fact, its value is actually greatest for $\omega = 90^\circ$
because
full phase is reached at periastron.  Due to the increased stellar heating
at this phase and viewing angle, water does not condense, but such condensation
does have a small effect as the orbit progresses toward greatest elongation.  Clearly,
the functional dependence of an EGP light curve on observer viewing angle and
cloud condensation can be quite complex.

\begin{figure}
\plotone{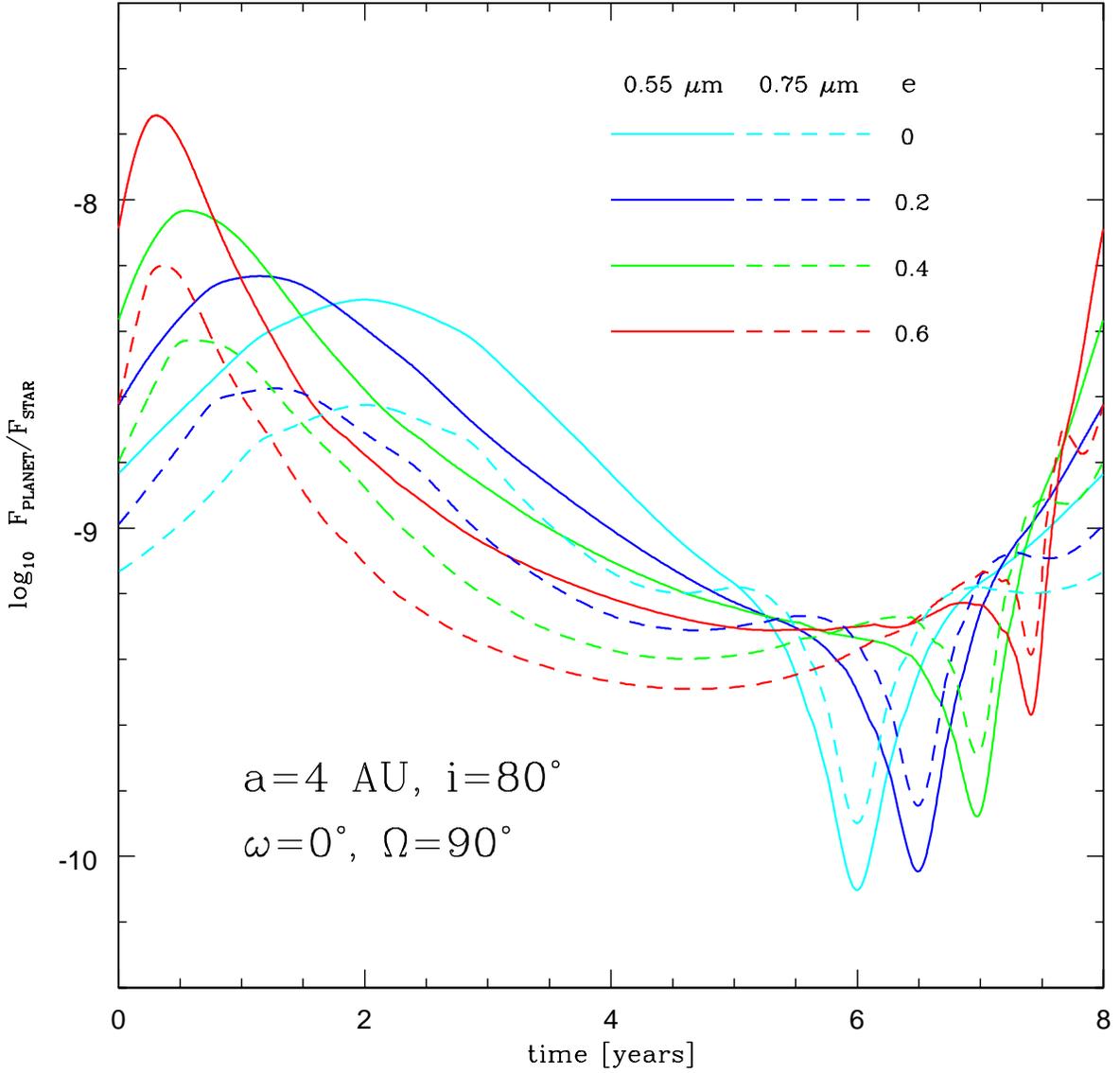} 
\caption{The logarithm of the optical (at 0.55 $\mu$m) and far red (0.75 $\mu$m)
planet/star flux ratios as a function of eccentricity for $a$ = 4 AU, fixing
$i$ at 80$^\circ$, $\Omega$ at $90^\circ$ and $\omega$ at $0^\circ$.  The planet/star
flux ratio is a factor of 2 to 3 greater at 0.55 $\mu$m than in the far red
at most planetary phases.
\label{fig_eccentricity2}}
\end{figure}

An EGP in a near-circular 4-AU orbit about a G2V star
contains a tropospheric water cloud deck.  With increasing eccentricity, ammonia
condenses above the water cloud deck when the planet is near apastron, resulting
in a two-cloud atmosphere.  As the planet orbits back toward periastron, the ammonia
cloud sublimates, leaving only the water cloud layer.
Figure \ref{fig_eccentricity2} depicts planet/star flux ratios (note log
scale) as a
function of eccentricity for $a$ = 4 AU, including the effects of cloud
condensation and fixing $i$ at $80^\circ$, $\Omega$ at $90^\circ$,
and $\omega$ at $0^\circ$.  We plot these flux ratios in the optical
(0.55 $\mu$m) and at 0.75 $\mu$m.  Although the 0.75 $\mu$m region has the
highest-albedo beyond 0.7 $\mu$m for most EGPs, the planet/star
flux ratio is still a factor of 2 to 3 below that in the optical region at
most planetary phases, which
indicates that the near-IR may be a more difficult spectral region
in which to detect reflected-light EGPs.  The periodic variation in
atmospheric composition (i.e. the appearance and disappearance of 
the ammonia cloud layer above the water cloud layer) has a less extreme
effect on the light curve than does the variation between cloudy and
cloud-free for the $a$=1.5 AU EGP discussed above.

\subsection{Effects of Inclination and Longitude of Ascending Node}
\label{sec_inclination}
\begin{figure}
\plotone{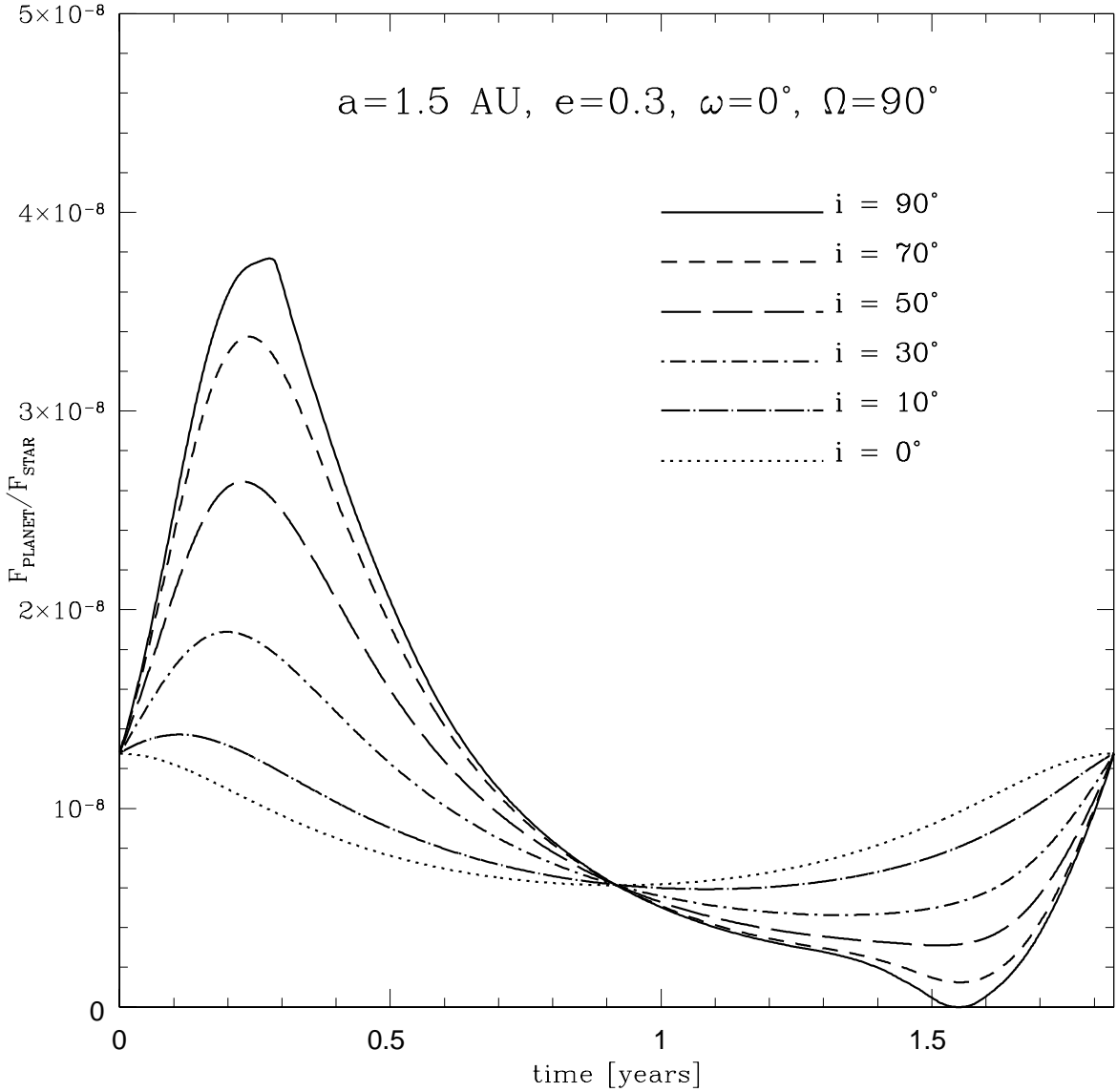} 
\caption{Variation with inclination of the optical light curve for an elliptical orbit
(G2V central star, $a$ = 1.5 AU, $e$ = 0.3).  For a highly-inclined orbit, the
peak of the planet/star flux ratio is a factor of $\sim$3 greater than for a face-on
($i=0^{\circ}$) orbit.  The (symmetric) variation for the face-on case is due entirely to the
variation in the planet-star distance over an eccentric orbit.
\label{fig_inclination}}
\end{figure}

By fixing the inclination of each orbit at 80$^\circ$ or 90$^\circ$, we have been
showing upper limits to the variation in planet/star flux ratios.  In Fig.
\ref{fig_inclination}, we illustrate how the light curve of an elliptical orbit
varies with inclination, given $\Omega$ = 90$^\circ$.  The optical peak
planet/star flux ratio of an EGP with $a$ = 1.5 AU
and $e$ = 0.3 orbiting a G2V star differs
by as much as a factor of 3 over the full range of $i=0^\circ$ to $i=90^\circ$.
There is variation in the light curve even for the face-on $i=0^\circ$ case,
an effect that is due entirely to the change in the planet-star distance
for this eccentric orbit.

\begin{figure}
\plotone{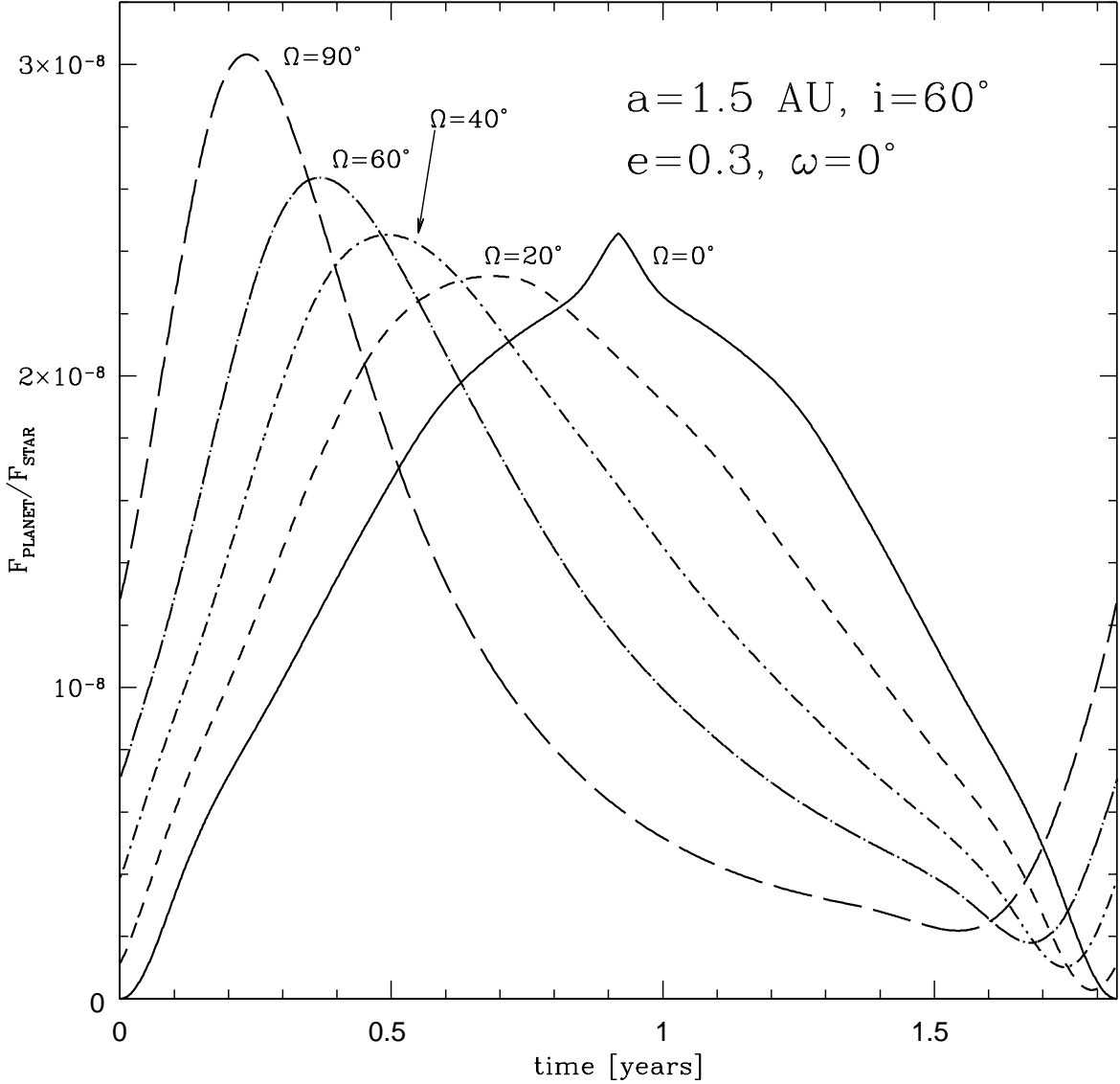}
\caption{Variation with the longitude of the ascending node ($\Omega$) of the optical
light curve for an elliptical orbit, assuming $a$ = 1.5 AU, $e$ = 0.3,
$i$ = 60$^\circ$, and $\omega$ = 0$^\circ$.  The peak of the light curve shifts from $\sim$0.25 years for
$\Omega$ = 90$^\circ$ to $\sim$0.9 years (half the orbital period) for $\Omega$ = 0$^\circ$.
The peak for $\Omega$ = 0$^\circ$ is the full-phase opposition effect.
\label{fig_bigomega}}
\end{figure}

We explore the effect on the light curve of varying the value of the
longitude of the ascending node from its default value of 90$^\circ$.
Recall that this parameter is the angle between the observer's line of sight and the
line formed by the intersection of
the observer's plane and the orbital plane (the line of nodes).  As this angle
decreases from 90$^\circ$ to 0$^\circ$ (or increases from 90$^\circ$ to 180$^\circ$),
the EGP orbit becomes edge-on, irrespective
of the orbital inclination, because the line of nodes becomes parallel to the observer's
line of sight.  Figure \ref{fig_bigomega} shows the optical
light curve for several different values of $\Omega$ for our $a$ = 1.5 AU, $e$ = 0.3,
$i$ = 60$^\circ$ EGP. 
The maximum of the light curve shifts from $\sim$0.25 years for
$\Omega$ = 90$^\circ$ to $\sim$0.9 years (half the orbital period) for $\Omega$ = 0$^\circ$.
For this particular set of parameters, the maximum planet/star flux ratio also decreases with $\Omega$,
until $\Omega$ = 0$^\circ$ is approached, where the opposition effect produces
a distinct peak at full phase.

\section{Summary and Prospects for Detection}
\label{sec_summary}
Our suite of model EGP phase functions and light curves from 0.2 AU to 15 AU
covers a large variation in atmospheric composition, from deep silicate cloud
and cloud-free EGPs with alkali metals, to giant planets with clouds of frozen ammonia
particles.  Self-consistent, wavelength-dependent modeling of EGP phase functions, albedos
and light curves reveals that:
\begin{itemize}
\item Albedos and phase integrals of EGPs are strongly
wavelength-dependent.  Most EGPs reflect incident
light to a larger degree in the optical than in the red or near-IR.
\item Planetary phase functions are wavelength-dependent.  Such differences
with wavelength result in color differences for a given EGP throughout the various
phases of its orbit.  In $V-R$ and $R-I$, cloud-free EGPs are bluest near greatest elongation,
while cloudy (water and/or ammonia) EGPs tend to be bluest in a gibbous phase.
\item Cloud-free EGPs exhibit smooth phase functions
and light curves with no significant ``opposition
effect.''  Cloudy EGPs may or may not exhibit an opposition effect, depending
on cloud particle sizes.  Only for nearly edge-on orbits
would such an effect be seen.
\item Small cloud particle sizes ($\sim$1 $\mu$m) produce higher planet/star
flux ratios than large particle sizes ($\sim$100 $\mu$m) at most optical and
near-IR wavelengths.  Furthermore, the shapes of cloudy EGP light curves depend
on both particle size and wavelength. 
\item Assuming highly-inclined circular orbits, at optical wavelengths EGPs are
3 to 4 times brighter near full phase than near greatest elongation.
\item EGPs in elliptical orbits can undergo major atmospheric compositional
changes, which may have significant effects on their light curves.  Additionally,
elliptical orbits generally introduce an offset between the time of the light
curve peak and the time of full planetary phase.
\item Because most wide-separation EGPs are in elliptical orbits, the Keplerian elements,
such as inclination, argument of periastron, and longitude of the ascending node, play
important roles in the shapes of their light curves.
\item The previous ``phase averaged'' planet/star flux ratios of Sudarsky,
Burrows, \& Hubeny (2003) and Burrows, Sudarsky, \& Hubeny (2004), derived with
a 1-D atmosphere code, are accurate to within 3\% in most cases.
\end{itemize}

The direct imaging of wide-separation EGPs at optical wavelengths, a lofty goal
by any measure, may become possible within the next several years by high contrast, space-based
imaging instruments such as {\it Eclipse} (Trauger \etal 2000, 2001) or
{\it EPIC} (Clampin \etal 2002; Lyon \etal 2003).
{\it Eclipse} is a 1.8-meter optical and near-infrared telescope
with an adaptive optics system.  High contrast imaging
of $\sim$10$^9$ in the optical, given an angular separation of 0.3$^{\prime\prime}$-2$^{\prime\prime}$,
may be possible.  If such sensitivity materializes, then a number of EGPs may be
detectable.  For example, an EGP in a 4-AU orbit about a G2V star at 10 parsecs
has an angular separation as large as 0.4$^{\prime\prime}$, and just beyond greatest elongation
(in a gibbous phase) it is expected
to exhibit an optical planet/star contrast above $\sim$2$\times10^{-9}$ (see
Fig. \ref{fig_lightcurves124}).
Even at 6 AU, and in this same phase, the planet-star contrast in the
optical is expected to be roughly $10^{-9}$.
Detailed models of specific EGP systems in combination with reliable ephemerides
are necessary both in the selection of good targets and for the physical
interpretation of positive observational results.

\acknowledgments

The authors are happy to thank Bill Hubbard, Jonathan Lunine,
Christopher Sharp, and Drew Milsom
for insightful conversations and help during the
course of this work, as well as
NASA for its financial support via grants NAG5-10760 and NNG04GL22G.
Furthermore, we acknowledge support through the Cooperative Agreement
NNA04CC07A between the University of Arizona/NOAO LAPLACE node and
NASA's Astrobiology Institute. 

\appendix
\section{Appendix} 
\label{append}
We solve the radiative transfer equation,
\begin{equation}
\label{rte}
\mu \frac{dI(\nu,\mu,\phi)}{dz} = -\chi(\nu) [I(\nu,\mu,\phi) - S(\nu,\mu,\phi)] \, ,
\end{equation}
where $I$ is the specific intensity of radiation at frequency $\nu$, and the direction
is specified by $\theta$ and $\phi$, where $\theta$ is the angle with respect to the normal to the surface
($\mu = \cos\theta$), and $\phi$ is the azimuthal angle. Furthermore, $z$ is
the geometrical coordinate, $\chi$ is the total extinction coefficient, and $S$ the source function,
which in the present case is given by
\begin{equation}
\label{sf1}
S(\nu,\mu,\phi)  = \frac{1-\epsilon_\nu}{4\pi}\,\int_{-1}^1d\mu^\prime 
\int_0^{2\pi}d\phi^\prime\, 
I(\nu,\mu^\prime,\phi^\prime) g(\nu,\mu^\prime,\phi^\prime,\mu,\phi) + 
\epsilon_\nu B_\nu\, ,
\end{equation}
where $1-\epsilon_\nu$ is the single-scattering albedo and $\epsilon_\nu$ is the photon
destruction coefficient, given by
\begin{equation}
\epsilon_\nu = \frac{\kappa_\nu}{\chi_\nu} \equiv 
\frac{\kappa_\nu}{\sigma_\nu + \kappa_\nu}\, ,
\end{equation}
where $\kappa_\nu$ is the coefficient of true absorption, and $\sigma_\nu$ the scattering coefficient.
Function $g(\nu,\mu^\prime,\phi^\prime,\mu,\phi)$ describes
a change of direction of a scattered photon, and is called the phase
function, or the single-scattering phase function.

The transfer equation (\ref{rte}) does not involve a coupling of the individual frequencies;
a solution may then be done frequency by frequency.  In the following text, we omit an
explicit indication of the frequency dependence. 
Introducing the optical depth, 
\begin{equation}
d\tau = -\chi dz\, ,
\end{equation}
we rewrite the transfer equation (\ref{rte}) in the usual form,
\begin{equation}
\label{rte2}
\mu \frac{dI(\mu,\phi)}{d\tau} = I(\mu,\phi) - S(\mu,\phi) \, .
\end{equation}
For a 
numerical solution of eq. (\ref{rte}) with the angle-dependent source function 
eq. (\ref{sf1}), we discretize the optical depth and both angles,
$\{\tau_d, d=1,\ldots, ND\}$, $\{\mu_i, i=1,\ldots,NMU\}$, 
and $\{\phi_j, j=1,\ldots,NPHI\}$,
and replace the integral in the source function eq. (\ref{sf1}) by a quadrature sum,
\begin{equation}
\label{sf2}
S_{i,j} = (1-\epsilon) \sum_{k=1}^{NMU} 
\sum_{l=1}^{NPHI} I_{k,l} g_{k,l,i,j} w^\mu_k w^\phi_l\, ,
\end{equation}
where $I_{k,l} \equiv I(\mu_k, \phi_l)$, and
$w^\mu$ and $w^\phi$ are the quadrature weights for the integrals over
$\mu$ and $\phi$, respectively.

The transfer equation (\ref{rte2}) is supplemented by the boundary
conditions. 
At the lower boundary, we assume the diffusion approximation,
\begin{equation}
\label{lbc}
I(\mu, \phi) = B + \mu \frac{dB}{d\tau}\, , \quad{\rm for}
  \quad \mu > 0\, ,
\end{equation}
where $B$ is the Planck specific intensity.  At the surface we have
\begin{equation}
\label{ubc}
I(\mu, \phi) = I^{\rm ext}(\mu, \phi)\, , \quad{\rm for}\quad \mu <0\, ,
\end{equation}
where $I^{\rm ext}$ is the specific intensity of the external radiation.
In the case of a point source, the external intensity is given by
\begin{equation}
I^{\rm ext}(\mu, \phi) = I^0 \delta(\mu - \mu^0) \delta(\phi)\, ,
\end{equation}
where $\delta$ is the Dirac $\delta$-function. Thus, we consider the external 
source at $\mu = \mu^0$ and $\phi=0$.

There are two different approaches to solve equation (\ref{rte2})
with the source function eq.~(\ref{sf2}): a direct one, and an
iterative one.  Sudarsky, Burrows, \& Pinto (2000) considered an azimuthally-averaged
specific intensity and redistribution function so that their specific
intensity depended on only one angle, $I=I(\mu )$ and 
$g=g(\mu ^{\prime },\mu )$. 
They solved eq.~(\ref{rte}) with an appropriately modified eq.~(\ref{sf1}) using
the Feautrier method (e.g., Mihalas 1978). This direct method solves for the     
angular coupling directly, but at the expense of inverting an $NMU\times NMU$  
matrix at each discretized depth point.

Generalizing a direct scheme to handle a dependence on two angles would be
very cumbersome and computationally costly, so we have developed another, much
faster, scheme based on the Accelerated Lambda Iteration method (ALI). 
Let us first take the simpler
case of isotropic scattering. In this case, the source function is given by 
\begin{equation}
S=(1-\epsilon)J+\epsilon B\, ,
\end{equation}
where 
\begin{equation}
J\equiv \frac{1}{4\pi}
\int_{-1}^{1}\int_{0}^{2\pi }\,I(\mu ,\phi )\,d\mu \,d\phi
\end{equation}
is the mean
intensity of radiation. The solution of the transfer equation may be written
as
\begin{equation}
J=\Lambda [S]\, ,
\end{equation}
where $\Lambda$ is an operator acting on the source function. Using this
relation, we obtain a single operator (integral) equation for the source
function, 
\begin{equation}
S=(1-\epsilon )\,\Lambda [S]+\epsilon B\, .  
\label{sint}
\end{equation}
The idea of accelerating the Lambda iteration consists of writing
\begin{equation}
\Lambda =\Lambda^{*}+(\Lambda -\Lambda^{*})\,,  \label{alidef}
\end{equation}
where $\Lambda^{*}$ is an appropriately chosen {\it approximate lambda
operator}. The iteration scheme for solving Eq.~(\ref{sint}) may then be
written as 
\begin{equation}
S^{(n+1)}=(1-\epsilon )\Lambda^{*}[S^{(n+1)}]~+~(1-\epsilon )(\Lambda
-\Lambda^{*})[S^{(n)}]~+~\epsilon B\,.  \label{snew1}
\end{equation}
The action of the exact $\Lambda$ operator is thus split into two
components: an approximate $\Lambda^{*}$ operator acting on the {\em new}
iterate of the source function, and the difference between the exact and
approximate operator, $\Lambda -\Lambda^{*}$, acting on the previous, {\em %
old}, and, thus, known source function. The latter contribution may be easily
evaluated by the formal solution. By the term ``formal solution'' we mean
a numerical solution of the transfer equation where the source function is
fully specified. 

Olson, Auer, \& Buchler (1986) showed that a nearly
optimum $\Lambda^{*}$ operator is the diagonal (local) part of the
exact $\Lambda$ operator, which can be easily evaluated. 
Inversion of $\Lambda^{*}$ is thus a simple algebraic
division. 

Applying the ALI idea to anisotropic scattering is not so straightforward
because the source function now depends on two angles, $\theta $ and $\phi $.
Thus, we introduce a ratio, $a_{\mu \phi }$, of the true scattering
term of the source function
to the angle-averaged one,
\begin{equation}
a_{\mu\phi} = \frac{\int_{-1}^1d\mu^\prime 
\int_0^{2\pi}d\phi^\prime\, 
I(\mu^\prime,\phi^\prime) g(\mu^\prime,\phi^\prime,\mu,\phi)}
{4\pi J}\, ,
\end{equation}
and use a scheme that proceeds as two
nested iteration loops:

1) estimate $a_{\mu\phi}$ (typically, initialize $a_{\mu\phi}=1$ );

2) while holding $a_{\mu \phi }$ fixed, iterate for $S$ exactly as
in the usual ALI treatment of the case of isotropic scattering;

3) after the inner loop is finished, update $a_{\mu\phi}$, and repeat.

\smallskip

The problem is thus essentially reduced to a set of formal solutions of the
transfer equation along individual rays defined by angles, $\theta $ and $%
\phi $. For the formal solution, we may use either the short characteristics
method (Olson \& Kunasz 1987; Hubeny 2003), or a Discontinuous Finite-Element
(DFE) method (Castor, Dykema, \& Klein 1992)

\smallskip

The single-scattering phase
function, $g(\Theta$), $\Theta$ being the scattering angle, is computed in
discrete values of $\Theta = \Theta_1, \ldots, \Theta_{NT}$, with
$\Theta_1=0$ and $\Theta_{NT}=\pi$.
However, in many cases the phase function is a very strongly peaked function of
$\Theta$, with a peak at $\Theta=0$ (forward scattering). 
Any simple angular quadrature thus yields inaccurate results, essentially 
because $g(\Theta_1=0)$ may be several orders of magnitude larger than
$g(\Theta_2)$, even for small values of $\Theta_2$ (in our calculations, we 
used $\Theta_2 = 1^\circ$). Because it is impractical to set up too may points very close
to $\Theta=0$, we devised the following procedure.
We split the phase function into two components:
one ($g'$) that has the same value at $\Theta_1=0$ as at $\Theta_2$
(or given by some suitable extrapolation from the values at $\Theta_3, \Theta_4$, etc.), 
and the second being the Dirac delta-function 
$\delta(\Theta)$, viz.
\begin{equation}
g(\Theta) = g'(\Theta) + \alpha \delta(\Theta)\, .
\end{equation}
This function is properly normalized to unity,
\begin{equation}
\int g(\Theta) d \Theta = \int g'(\Theta) d \Theta + \alpha = 1\, .
\end{equation}
Now, with this form of the phase function, one can analytically handle
the source function. The scattering contribution to the source function
becomes (schematically):
\begin{eqnarray}
S^{\rm sct}(\mu, \phi) =  \frac{1}{4\pi}\int_{-1}^1d\mu^\prime \int_0^{2\pi}d\phi^\prime\,
            I(\mu', \phi')  g(\mu', \phi', \mu, \phi) \\
           = \int_{-1}^1d\mu^\prime \int_0^{2\pi}d\phi^\prime\,
              I(\mu', \phi') g'(\mu', \phi', \mu, \phi) + 
             \alpha I(\mu, \phi)
\end{eqnarray}
and, thus, the term $\alpha\, I(\mu, \phi)$ can be understood as a negative contribution
to the absorption coefficient.  Therefore, one can use, in the angle-dependent
transfer calculations, the modified phase function $g'$, which is smooth
and well-behaved, and the scattering contribution to the absorption
coefficient $\chi$ will be modified to $(1-\alpha)\, \chi$. In other words,
very strong forward-scattering effectively reduces the optical 
thickness of the cloud.

{}

\clearpage

\clearpage
\begin{figure}
\plotone{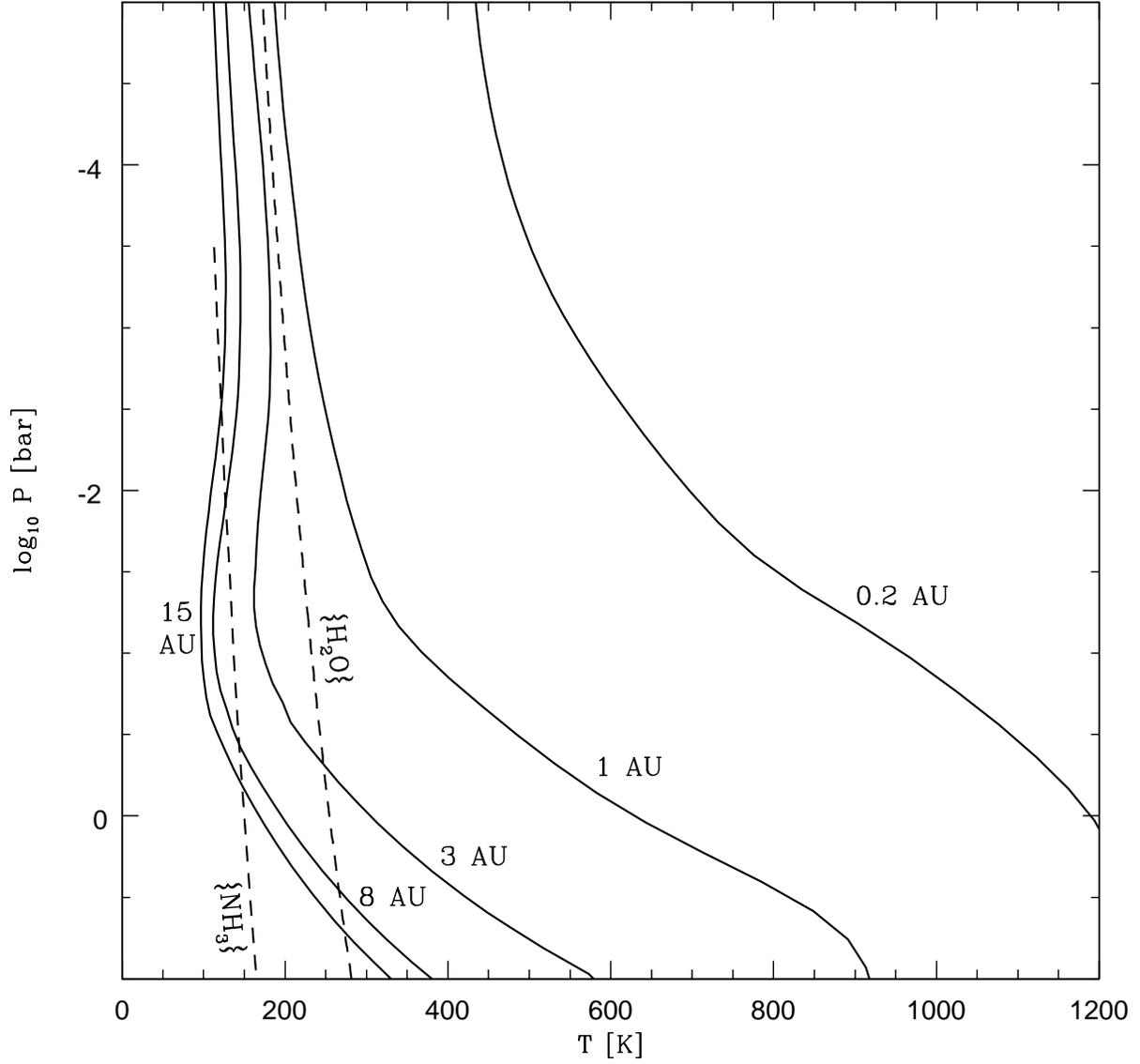}
\caption{The temperature-pressure (T-P) profiles for a selection of
our model EGPs.  Condensation curves for water and ammonia are shown,
while those for forsterite and iron are off the scale, at higher
temperatures.  The deeper intersections of these condensation curves with
the T-P profiles indicate the positions of the cloud bases.
\label{fig_TPprofiles}}
\end{figure}

\clearpage

\clearpage

\clearpage

\clearpage

\clearpage

\clearpage

\clearpage

\clearpage

\clearpage

\clearpage

\clearpage

\clearpage

\clearpage

\clearpage

\clearpage

\clearpage

\clearpage

\clearpage

\clearpage

\clearpage

\end{document}